\documentstyle[epsf]{mn}
\newcommand{\etal}{{et al}\/.}
\begin{document}
\title[{\it ROSAT} X-ray observations of 3CRR radio sources]{{\it ROSAT}
X-ray observations of 3CRR radio sources}
\author[M.J.~Hardcastle \& D.M. Worrall]{M.J.\ Hardcastle$^1$ and
D.M.\ Worrall$^{1,2}$ \\
$^1$Department of Physics, University of Bristol, Tyndall Avenue,
Bristol BS8 1TL\\
$^2$Harvard-Smithsonian Center for Astrophysics, 60 Garden Street,
Cambridge, MA 02138, U.S.A.}
\maketitle
\begin{abstract}
Over half the 3CRR sample of radio galaxies and quasars has been
observed in X-rays with {\it ROSAT} pointed observations, and we
present results from these observations, discussing many of the
sources in detail. The improved spatial resolution of {\it ROSAT} over
earlier missions allows a better separation of the nuclear and
extended components of the X-ray emission. We investigate the
relationship between nuclear X-ray and core radio luminosity and show
that our results support a model in which every radio galaxy and
quasar has a beamed nuclear soft X-ray component directly related to
the radio core. We report evidence for rich cluster environments
around several powerful quasars. These X-ray environments are
comparable to those of high-redshift radio galaxies.
\end{abstract}
 
\section{Introduction}
\label{intro}

Soft X-ray emission from radio galaxies and radio-loud quasars probes
a range of physical conditions. On the largest scales, the X-rays are
the thermal emission of hot intracluster gas. In general we do not
expect a strong quantitative relationship between this emission and
the radio properties of the galaxy or quasar, although there is direct
evidence for the influence of one on the other in a few well-studied
objects [e.g. NGC1275, B\"ohringer \etal\ (1993); Cygnus A, Carilli,
Perley \& Harris (1994); 3C\,449, Hardcastle, Worrall \& Birkinshaw
(1998c)].

On smaller scales, X-ray emission may come directly from components of
the radio source, as synchrotron, inverse-Compton, or synchrotron
self-Compton (SSC) radiation. There is evidence for X-ray synchrotron
emission in the jets of M87 (Biretta, Stern \& Harris 1991) and
Centaurus~A (Turner \etal\ 1997) and in a hotspot of 3C\,390.3 (Prieto
1997), and for a synchrotron self-Compton process in the hotspots of
Cygnus A (Harris, Carilli \& Perley 1994). Moreover, inverse-Compton
scattering of CMB or active nucleus photons may be responsible for
extended emission seen in Fornax A (Feigelson \etal\ 1995) and 3C\,219
(Brunetti \etal\ 1998). However, in general such kpc-scale X-ray
emission is likely to be weak compared to that due to hot gas.

Inverse-Compton processes are more important on parsec scales, where
electron and photon densities are much higher. Such emission is
unresolved by present-day X-ray observatories. Synchrotron emission,
along with thermal emission from the central engine and accretion
disc, may also contribute to an unresolved nuclear
component. Obscuration has a strong effect on what is
observed. Unified models of radio sources (e.g. Urry \& Padovani 1995)
require an obscuring `torus' of dusty material around the central
engine, so that nuclear optical continuum and broad lines are not seen
in objects classed as radio galaxies.  This torus should also obscure
soft X-ray emission originating close to the central engine, leading
to suggestions (e.g.\ Crawford \& Fabian 1996a) that soft X-ray
emission from powerful radio galaxies should in general be dominated
by the extended thermal component.  In hard X-rays, a strongly
absorbed nuclear component has been found in a few radio galaxies,
consistent with this picture (e.g. Ueno \etal\ 1994).

However, the radio-related component of the nuclear emission may
originate on scales larger than those of the torus. Although not
affected by obscuration, it would be rendered anisotropic by bulk
relativistic beaming. There is substantial evidence that unabsorbed
radio-related non-thermal X-rays are seen in radio galaxies.  Recent
work has strengthened the conclusion of Fabbiano \etal\ (1984) that
the correlation between nuclear soft X-ray and radio-core luminosity
implies a jet-related origin for the X-ray emission, as
high-resolution X-ray observations have allowed point-like and
extended components to be separated (Worrall \& Birkinshaw 1994; Edge
\& R\"ottgering 1995; Worrall 1997). A relativistically beamed
radio-related component almost certainly dominates the soft X-ray
emission in core-dominated quasars and BL Lac objects.

A large, unbiased sample of objects is most effective for
investigating the origins of X-ray emission in extragalactic radio
sources.  The 3CRR sample (Laing, Riley \& Longair
1983; Laing \& Riley, in preparation) is a flux-limited sample of the
brightest radio galaxies and quasars in the Northern sky. It includes
all objects with 178-MHz flux density greater than 10.9 Jy (on the
scale of Baars \etal\ 1977) having $\delta > 10^\circ$ and $|b| >
10^\circ$.  Excluding the starburst galaxy 3C\,231 (M82), which will
not be considered further in this paper, it contains 172 objects, of
which, at the time of writing, 89 have been observed (either
intentionally or serendipitously) in {\it ROSAT} pointed observations
at off-axis angles less than 30 arcmin (and usually on-axis).
Most sources in the sample are powerful, distant FRII
(Fanaroff \& Riley 1974) radio galaxies and quasars, and almost all 
are well-studied in the radio and optical, making 3CRR an ideal sample
for a comparison of radio and X-ray properties. Although such
comparisons have been carried out previously (e.g.\ Fabbiano \etal\
1984; Crawford \& Fabian 1995, 1996a; Prieto 1996), none has made full
use of {\it ROSAT}'s high spatial resolution to try to distinguish
between different physical origins for the X-ray emitting components,
and in particular to test the suggested relationship between X-ray and
radio core emission.

In this paper we tabulate the X-ray count rate from each of the 89
{\it ROSAT}-observed 3CRR sources and present our best estimates of
the count rate associated with a central unresolved component. We use
radio data from the literature to discuss whether a coherent picture
of the physics can be derived from the X-ray and radio
observations. Complex sources are discussed in detail and we provide
references to existing studies. Some objects are the subject of
more detailed work which will appear elsewhere.

Throughout the paper we use a cosmology in which $H_0 = 50$ km
s$^{-1}$ Mpc$^{-1}$ and $q_0 = 0$. Spectral index is defined in the
sense that flux density is proportional to $\nu^{-\alpha}$.

\section{Data and analysis}
\label{analysis}

Table \ref{sources} lists the 89 {\it ROSAT}-observed sources. [The
{\it ROSAT} Master observation catalogue lists a further three 3CRR
objects (3C\,34, 3C\,336 and 3C\,438) for which data were not public
in time for our study. With the demise of {\it ROSAT}, further data
will not be forthcoming.] Optical and radio classifications and
redshifts come largely from the updated 3CRR tables of Laing \& Riley
(in prep.), updated with more recent spectroscopic observations where
available (e.g.\ Laing \etal\ 1994; Laing, private communication;
Jackson \& Rawlings 1997). The classification of objects by whether or
not they have strong high-excitation nuclear emission lines dates back
to Hine \& Longair (1979); we use the definition of Laing \etal\
(1994) who define high-excitation objects as having [OIII]/H$\alpha >
0.2$ and equivalent widths of [OIII] $> 3$\AA.  Low-excitation objects
are excluded from many radio galaxy-quasar/broad line radio galaxy
unified models (e.g. Barthel 1994) because broad lines in their
spectra would be very hard to detect, so that there is no optical
evidence for their orientation to the line of sight. They also appear
to have rather different radio properties from high-excitation radio
galaxies (Jackson \& Rawlings 1997, Hardcastle \etal\ 1998b).

For each source we give the {\it ROSAT} observation request (ROR)
number of the primary data set we have used and the source's angular
offset from the pointing direction where this is significantly
different from zero. Because we are most interested in the sources'
spatial properties in this paper, we have in general preferred HRI
observations over PSPC where both exist (the 50 per cent encircled
energy radii are $\sim 13$ and $\sim 3$ arcsec for the PSPC at 1 keV
and the HRI, respectively), though we have checked all HRI results against
any existing PSPC data for consistency. Where an X-ray source is
reasonably bright, and there is no obvious evidence for significant
source extension, we have used an on-source circle of radius $\sim 1$
arcmin and taken background from an annulus of radii 1 and 2 arcmin
about the source (for HRI data) or $\sim 2$ arcmin and 2 and 3 arcmin
(for PSPC data), taking care in all cases to exclude unrelated
contaminating sources. Where a smaller source circle was used, we have
corrected the derived count rate to that expected from the standard
source and background regions, for consistency (this affects only a
few sources). Counts and background-subtracted count rates for the few
sources observed significantly off-axis are corrected for the effects
of vignetting (calculated at 1 keV) and any additional
obscuration. Upper limits on the total (resolved plus unresolved)
counts, quoted where a source was undetected, are $3\sigma$ limits,
derived by applying Poisson statistics to a detection cell of a size
appropriate to the off-axis angle of the source [for on-axis sources,
the sizes used are $5 \times 5$ arcsec (HRI) and $30 \times 30$ arcsec
(PSPC)].

The well-known `aspect smearing' problem of the HRI (e.g.\ Morse 1994)
means that we cannot reliably distinguish unresolved from resolved
components in HRI data simply by fitting the nominal HRI point
response function (PRF), and determining the number of counts
associated with a possible central point source necessarily involves a
certain amount of interpretation. Our procedure, justified by detailed
work on sources known to be point-like, is that where the X-ray
emission from a source is obviously compact, we treat it all as
point-like if its radial profile is consistent with a nominal HRI PRF
convolved with a Gaussian of FWHM a few arcseconds. If the statistics
are too poor to make this judgement, we use the total measured counts
as an upper limit on the counts in a possible point-like component.

Where there is emission which is unambiguously extended we attempt to
separate out a compact, possibly AGN-associated, component. Our best approach, appropriate where the data are
roughly radially symmetrical, is to fit models convolved with the
instrumental PRF to the radial profile
of the emission. $\beta$-models (e.g. Sarazin 1986) are physically
appropriate for extended emission originating from hot gas in
hydrostatic equilibrium, and provide a useful way of characterising
extended emission even when the origins of the emission are not
known. In all such fits the parameter $\beta$ is allowed to take the
values 0.35, 0.5, 0.667 and 0.9, while core radius is varied over a
wide range, and the central normalisations of both the $\beta$ model
and an unresolved central source are allowed to vary, so that the
counts associated with the best-fitting components need not sum
exactly to the total counts obtained directly from the data. We
calculate counts in the best-fitting $\beta$ model from the integral
of the model over the source and background regions.

In radial-profile fitting we are hampered, in the case of HRI
data, by our inadequate knowledge of the PRF for a particular
observation because of aspect-smearing effects. Some techniques exist
to circumvent this problem. Where a bright nearby pointlike source is
available in the image, it is possible to use it as a template to
determine the PRF applicable to the observation (e.g.\ Hardcastle,
Lawrence \& Worrall 1998d), but such sources are rare. Another possible
technique is `dewobbling' the source (Harris \etal\ 1998b). The aspect
errors in HRI observations are thought to arise from pixel-to-pixel
gain variations in the aspect camera CCD. The
spacecraft wobble exacerbates this problem because it causes the aspect
stars to move across the CCD on short time-scales. The
`dewobble' technique relies on binning the data according to
the phase of the wobble, and then recentroiding to correct for aspect
errors; so long as the satellite roll angle and the properties of the
CCD are constant, the aspect error should be a function of the wobble
phase only. Harris \etal\ provide {\sc iraf}/PROS scripts to perform
this task. The main disadvantage of the method is that it is applicable only to
sources with a relatively bright unresolved component (of order 0.1
counts s$^{-1}$); this rules out its application to many of the
sources in the sample. 

In general we expect sources that have been dewobbled to show some
deviation from the nominal PRF on small scales; the procedure of
stacking by centroid should tend to produce a profile that is more
sharply peaked than the true PRF, as can easily be understood by
considering the limit of a single photon in each bin, but on the other
hand errors in centroiding and restacking will broaden the
profile. Where HRI observations are clearly dominated by a point-like
component (e.g.\ quasars) but may also contain some extended emission,
we sometimes use the central regions of the (dewobbled) radial profile
of the source itself to determine a broadened PRF for model fitting;
the procedure here is to fit the central part of the radial profile
with the nominal HRI PRF convolved with a Gaussian, assuming that the
central source is pointlike. All cases where dewobbling or a broadened
PRF have been applied are discussed individually in section
\ref{notes}. Where neither method is applicable we assume that the
nominal PRF (David \etal\ 1997) applies.

An error in the automatic (SASS) processing of HRI observations has
recently been discovered (Harris 1999). This introduces an additional
aspect uncertainty into HRI data taken before 1997 January 17. We have
applied the SAO-provided scripts which approximately correct this
problem to all affected HRI observations in which we have carried out
radial profile fits. In general the scripts make a small difference to
the radial profiles of our target sources, most noticeable in those
sources, mainly quasars, which are dominated by a central point-source
component. In no case does the correction cause a previously existing
extended component to disappear.

The PRF of the PSPC is sufficiently broad that it is not expected to
be significantly affected by aspect problems, and PSPC data are not
affected by the SASS processing error. When fitting models to
data taken with this instrument, we determine an energy-dependent PRF
from the data in the energy band 0.2--1.9 keV (where the PRF is well
known). For consistency of presentation, the results of model fitting
are scaled to the full 0.1--2.4 keV passband of {\it ROSAT} using an
appropriate spectral model.

\section{Notes on individual sources}
\label{notes}

In this section we comment on any features of particular interest in
the X-ray observations, and discuss our analysis of complicated
objects. Where we have separated a source into extended and compact
components, values for the best-fitting parameters of the
$\beta$-model and point source are listed in Table \ref{profile}.

As well as the core radius and $\beta$ for the best-fitting
$\beta$-model, we tabulate the central surface brightness $b_0$. This quantity may be converted to an estimate
of the central proton density of the X-ray emitting gas, $n_{p0}$. From
Birkinshaw \& Worrall (1993), eq. 10, to within a factor of 2
\[
n_{p0}^2 \approx 2 \times 10^{-9} {{(1+z)^6 b_0 f}\over{D_L \theta_c}}
\] 
where $b_0$ is in counts s$^{-1}$ arcsec$^{-2}$, $\theta_c$ is the
core radius in arcsec and $D_L$ is the luminosity distance to the
source in Gpc. $f$, the factor relating count rate to
distance-normalised emission measure. $f$ has a relatively weak
dependence on spectral shape, since the passband of {\it ROSAT} is
broad. Within a factor of about 1.5, for temperatures between 1 and 5
keV, $f$ has the value $4.8 \times 10^{22}$ m$^{-5}$ s count$^{-1}$
for HRI data and $1.5 \times 10^{22}$ m$^{-5}$ s count$^{-1}$ for the
PSPC. The central density can be used to calculate the central cooling
time, using the relation given by Sarazin (1986):
\[
t_{\rm cool} \approx 3 \times 10^{13} n_{p0}^{-1} \sqrt{kT}
\]
where $kT$ is in keV and $t_{\rm cool}$ in years.

Where we quote a central density or cooling time in the present paper,
we have calculated the value of $f$ appropriate for the source's
redshift, $N_H$ and estimated temperature.

\subsection{3C\,28}

This low-excitation radio galaxy lies in the double X-ray cluster
Abell 115 (Feretti \etal\ 1984; Worrall, Birkinshaw \& Cameron
1995). It has a relaxed double radio structure with no detected radio
core, although thin radio filaments emerging from the host galaxy may
suggest that it is not an inactive object (Leahy, Bridle \& Strom
1998). 3C\,28 lies in the northern component of the X-ray cluster,
with its lobes straddling (within the {\it ROSAT} pointing accuracy)
the peak in the X-ray surface brightness. Radial profile fitting shows
the source to be significantly extended (Table \ref{profile}). A
point-like component is not significant on an F-test. It seems likely
that most of the emission from the central peak originates in the
central regions of a cooling flow, as suggested by Feretti \etal\
(1984). We therefore take the best-fit point-like component as an
upper limit on the AGN contribution.

\subsection{3C\,31}

Our results for this FRI radio galaxy (NGC 383) are consistent with
Trussoni \etal\ (1997). PSPC images show extended emission, but the
HRI detects only point-like emission, with a component coincident with
3C\,31's nucleus. Emission coincident with NGC 380 and NGC 379, other
members of 3C\,31's group, is also detected.

\subsection{3C\,33}

Results for this low-redshift FRII radio galaxy are taken from Hardcastle,
Birkinshaw \& Worrall (1998a). The source is unresolved with the HRI.

\subsection{3C\,48}

The HRI data for this compact steep-spectrum quasar seem to show a
strong E-W elongation, but this is mainly contributed by a single
observation interval (OBI).  The observations of 3C\,48 consist of 7
OBIs taken over the course of three days at three distinct nominal
satellite roll angles. We first split the data into these three
`stable roll angle intervals' and then applied the dewobbling
procedure. The result was a clear reduction in the east-west
elongation of the data. The radial profile of the source was still not
well fit with a nominal HRI PRF; there were large contributions to the
fitting statistic both from the inner bins, on scales of a few
arcseconds, and from the bins on scales of 20--50 arcsec. We therefore
used the central part of the radial profile as a template for the
PRF. This left an excess component of extended X-ray emission, on
scales of tens of arcsec, which we interpret as emission from hot gas;
results are given in Table \ref{profile}. PSPC data are adequately
modelled as a point source but not inconsistent with the HRI-derived
model. For any reasonable temperature assumption, the bright, compact
extended X-ray emission implies rapid cooling with mass deposition
rates of several hundreds of solar masses per year.  This would tend
to support the cooling-flow model of Fabian \etal\ (1987) for the
optical emission-line regions. The X-ray emission from 3C\,48 is
discussed in more detail elsewhere (Worrall \etal , in preparation).

\subsection{3C\,61.1}

A faint point source is detected close to the pointing centre of the
HRI observation, but it is $\sim 30$ arcsec from the core of the FRII
radio galaxy (Laing \& Riley, in prep.), a larger offset than would be
expected if the sources were associated (see Fig.~\ref{61.1fig}). The
X-ray source appears to be almost coincident with a point-like optical
object in the field, about 30 arcsec to the east of 3C\,61.1's host
galaxy (see the plates of Gunn \etal\ 1981), which is
presumably a background active nucleus. We therefore derive an upper
limit on the X-ray counts associated with the radio position.

\subsection{3C\,84}

This source (Perseus A, NGC 1275) lies in the centre of the Perseus
cluster, and in a strong cooling flow. Remarkable shell-like structure
is apparent in the HRI image, which B\"ohringer \etal\ (1993) relate
to displacement of thermally emitting gas by the radio
lobes. Consequently it is hard to measure a good count rate for the
central component; radial fitting is not appropriate. The counts we
associate with the unresolved source are measured from a 15-arcsec
radius source circle and a 15--25-arcsec background annulus in the
centre, rather than by fitting to the data. For consistency with the
other objects, the counts we tabulate are corrected to our standard
1-arcmin source circle and 1--2-arcmin background region. The total
counts listed are for a 7-arcmin source circle and 7-8 arcmin
background region.

\subsection{3C\,98}

This nearby FRII galaxy shows some extended X-ray emission on scales
of tens of arcseconds, in addition to a compact central source (Table
\ref{profile}).

\subsection{3C\,123}

This peculiar FRII radio galaxy is dominated by an X-ray extended
component. Radial profiling does not require a core component, but the
X-ray distribution is complex. Since radial profiling may not be
appropriate, we derive an upper limit on the core contribution from a
15-arcsec source circle around the brightest X-ray peak. The counts we
tabulate are corrected to our standard 1-arcmin source circle and
1--2-arcmin background region. We will discuss this object in more
detail elsewhere (Hardcastle \etal , in preparation).

\subsection{3C\,212}

This quasar is unresolved with the PSPC. Our count rate is consistent
with that measured from the same data by Elvis \etal\ (1994).

\subsection{3C\,215}

This object is a well-studied lobe-dominated quasar. Both the co-added
HRI image used in our analysis and, to a lesser extent, the individual
datasets from which we generated it show elongation of the X-rays in a
NNW-SSE direction; the weak nearby point sources in the field also
show elongation roughly in this direction, so we attribute the
elongation to aspect-smearing problems. The long observation contains
many different OBIs and satellite roll angles, and to apply dewobbling
it was necessary to drop some of the shorter intervals with
insufficient counts for good centroiding, leaving 65 ks, or 75 per
cent of the available data. The restacked image is indeed narrower and
without significant elongation. Radial profile fits, using a PRF based
on the inner regions of the source, show the dewobbled source to have
a significant extended component (Table \ref{profile}). Galaxy counts
around 3C\,215 also suggest a rich cluster environment (Ellingson, Yee
\& Green 1991) and some extended line emission is seen (Crawford \&
Fabian 1989) so it seems likely that we are seeing emission from hot
cluster gas, perhaps influencing the distorted structure of the radio
source (Bridle \etal\ 1994). The core radius corresponds to 140 kpc
and the cooling time of the gas will be less than $10^{10}$ years for
$kT \la 4$ keV.

\subsection{3C\,219}

This broad-line radio galaxy (Laing \etal\ 1994) shows largely
point-like emission in the short archival PSPC observation, but there
is some evidence of excess counts in the radial profile on scales of
40--100 arcsec. This may be related to the poor optical cluster known
to be associated with 3C\,219 (Macdonald, Kenderdine \& Neville 1968),
though Brunetti \etal\ (1998), using HRI observations not yet in the
public archive, suggest instead that the extension is inverse-Compton
emission associated with the radio lobes, while not ruling out a
thermal origin for some or all of it. A point-source fit to the PSPC
radial profile has $\chi^2 = 17.3$ with 9 degrees of freedom, a barely
acceptable fit, while the addition of the $\beta$-model (Table
\ref{profile}) improves the fit significantly. The point-like
component dominates, contributing $75 \pm 7$ per cent of the counts in
the 2-arcmin source circle.

\subsection{3C\,220.1}

Our results for this FRII radio galaxy are taken from Hardcastle,
Lawrence \& Worrall (1998d). The source is decomposed into point-like
and extended components (Table \ref{profile}). The counts allocated to
the point source here neglect any possible contribution from a cooling
flow.

\subsection{3C\,220.3}

Our results for this undetected FRII radio galaxy are taken from
Worrall \etal\ (1994).

\subsection{4C\,73.08}

This low-excitation FRII source is in a crowded X-ray field. The
optical position is coincident with a weak extended X-ray source, but
there are too few counts for adequate radial profile fitting. We take
the counts in a 2-arcmin source circle as an upper limit to point-like
emission.

\subsection{3C\,254}

The radial profile of the PSPC observation of this quasar is poorly
fit by a point-source model, with positive residuals on scales between
30 and 60 arcsec. The HRI observation detects a weak secondary source
30 arcsec to the E, unidentified on Digitized Sky Survey
(DSS)\footnote{Available on the World Wide Web at URL:
$<$http://stdatu.stsci.edu/dss/$>$} plates, and too close to the
quasar to be resolved by the PSPC. The secondary source is not
responsible for all of the extension seen in either the HRI or PSPC
data; it may, in any case, be a clump in extended thermal
emission. There is optical evidence that 3C\,254 lies in a cluster of
galaxies (Bremer 1997), and there is no evidence in the HRI images for
X-ray emission from the foreground galaxy to the NE of the quasar reported
by Bremer (1997). We fit to the radial profile of the PSPC
observations, where there are more counts and so better statistics;
dewobbling has no effect on the results. The results are tabulated in
Table \ref{profile}. The counts in the extended component indicate a luminous
cluster; no cooling flow is required unless $kT\la 2.5$ keV, but we cannot
rule out a contribution to the point-source component from unresolved
cooling gas. If we assume $kT = 7.9$ keV, then the rest-frame 0.3--3.0
keV luminosity of the extended component is $10^{38}$ W, consistent
with the estimate of Crawford \& Vanderreist (1997) who examine the
spectrum of the same PSPC data. Fitting to the radial profile of the
dewobbled HRI data gives a smaller and somewhat less luminous
extended component, but the results are consistent within the
uncertainties on the fitted parameters.

\subsection{3C\,263}

After dewobbling the long HRI observation of this quasar, known to lie
in a cluster of Abell richness $\ga 1$, we find no significant
evidence of extended emission. This is consistent with the results of
an earlier study using the same data by Hall \etal\ (1995).

\subsection{3C\,263.1}

This FRII radio galaxy was not detected by Crawford \& Fabian (1996a)
in a far-off-axis PSPC observation. However, an HRI observation
detects a source within 6 arcsec of the optical position of 3C\,263.1,
with a count rate consistent with the upper limit of Crawford \&
Fabian. There is no obvious background source on the optical plates of
Laing \etal\ (1978) that might be responsible for the X-ray
emission. The HRI emission is compact and adequately fitted by a
broadened PRF. The source is surprisingly bright if it is due only to
nuclear X-ray emission, given the weak flux of the radio core (Liu,
Pooley \& Riley 1992). Up to $\sim 60$ per cent of the X-ray flux may
be contributed by a cluster of comparable size to that seen in
3C\,220.1, but the data do not require this, so for consistency we
treat the source as pointlike.

\subsection{3C\,264}

This narrow-angle-tail radio galaxy lies $\sim 7$ arcmin away from the
X-ray centroid of the cluster Abell 1367, and 8.4 arcmin off-axis in
the PSPC observation. We use the parametrised off-axis PSPC PSF
(Hasinger \etal\ 1995) to perform a radial-profile fit in a standard
2-arcmin source region, which excludes most of the emission from the
cluster. The results (Table \ref{profile}) suggest that there is some
extended emission associated with the radio source, in addition to a
strong point-like component. Our result is consistent with the count
rate quoted by Edge \& R\"ottgering (1995).

\subsection{3C\,270.1}

50 arcsec to the NW of this high-redshift quasar there is a second
compact source, associated with a faint compact source on DSS plates
and so probably another AGN. Our counts are taken from a 30-arcsec
source circle around the quasar, corrected to our standard 2-arcmin
source circle and 2--3-arcmin background annulus. We are unable to say
whether the apparent extended emission around 3C\,270.1 is related to
our target source or the other object.

\subsection{3C\,272.1}

This nearby FRI source (M84) is extended in the X-ray. Our results are
taken from a fit of point source plus $\beta$ model (Table \ref{profile}).

\subsection{Abell 1552}

The cluster is detected in the short PSPC observation, but no
point-like excess is found at the optical position of the radio galaxy.

\subsection{3C\,274 (M87)}

Harris, Biretta \& Junor (1997) present evidence that the X-ray core
of M87 varies by $\sim 20$ per cent on time-scales of years. We use
their core count rate for June 1995. The total counts tabulated are
for the source region they define, a 276-arcsec radius circle with
background taken from a 280--300 arcsec annulus, both centred around
the X-ray core, and our results are consistent with theirs.

\subsection{3C\,275.1}

Both HRI and PSPC data are available for this quasar. The PSPC data
are confused by weak extended emission from the nearby spiral galaxy
NGC 4651; when this is removed, the data show a point-like source. The
longer HRI observation shows a profile slightly broader than the
nominal PRF, characteristic of a source affected by aspect smearing.
When the profile is fitted with a broadened PRF, there is some
indication of additional extension on scales of 20--60 arcsec.  Using
the broadened PSF, we fit an additional $\beta$-model component, with
results tabulated in Table \ref{profile}. As in 3C\,215, the galaxy
counts of Ellingson \etal\ (1991) imply a rich cluster environment for
this source, so it is likely that the extended emission we detect is
real and due to hot intracluster gas. Hinten \& Romanishin (1986)
suggest that the extended emission-line region around the source is
due to a cooling flow, but the best-fitting $\beta$-model does not
require cooling.

\subsection{3C\,277.2}

Our count rate for this faint FRII radio galaxy (taken from a
30-arcsec source circle and corrected to our standard 1-arcmin source
region) is consistent with that of Crawford \& Fabian (1995).

\subsection{3C\,280}

Worrall \etal\ (1994) describe the separation of the PSPC emission
from this high-redshift FRII radio galaxy into compact and extended
components. The archival HRI data are consistent with such a
decomposition (cf. Dickinson \etal\ 1998), and so we use the Worrall
\etal\ results here (Table \ref{profile}).

\subsection{3C\,288}

Because this radio galaxy is 21 arcmin off-axis in the PSPC
observation, uncertainties in the PSPC PRF at this radius mean that we
have not been able to separate extended and compact emission, although
the source is at least partially extended. 3C\,288 appears in the
optical to be the dominant member of a cluster (Wyndham 1966) and so
it is not surprising that extended X-ray structure is
seen. Interactions with the cluster gas may be the reason for its
unusual structure in the radio (Bridle \etal\ 1989); it is one of a
small class of reasonably powerful, distorted double sources with
strong jets but weak hotspots, which often have a low-excitation
optical spectrum and an association with a cluster [other good
examples are 3C\,401 and 3C\,438, Hardcastle \etal\ (1997); the class
is similar in radio properties to the `jetted doubles' of Law-Green
\etal\ (1995)]. The available X-ray data are unfortunately not good
enough to test whether 3C\,288 lies in a cooling flow, a suggested
origin for this type of source.
\label{jetted}
\subsection{3C\,294}

This high-redshift radio galaxy was detected in PSPC observations by
Crawford \& Fabian (1996a). The counts we measure from a long
archival HRI observation are consistent with
the result of Crawford \& Fabian within the errors. The source appears
extended, but the low count rate prevents us from extracting a
reliable radial profile. Dickinson \etal\ (1998) reach similar conclusions.

\subsection{3C\,295}

This radio galaxy is known to lie in an optical cluster (e.g.\ Yates,
Miller \& Peacock 1989) and to have extended X-ray emission (Henry \&
Hendriksen 1986). Our best-fitting characterisation of the HRI
emission with a $\beta$-model and point source is tabulated in Table
\ref{profile}. Acceptable fits are found with no point-source
contribution. Our values for core radius and $\beta$ are consistent
with the findings of Henry \& Hendriksen. The point-source
contribution is much lower than their adopted value (which would
correspond to approximately 115 counts in the HRI observation, on the
assumption of a power-law spectrum and galactic absorption) but agrees
within the errors. Our best-fitting $\beta$ model is also consistent
with Neumann (1999). The best-fitting model implies rapid cooling,
with a mass deposition rate of some hundreds of solar masses per year,
consistent with the picture of Henry \& Hendriksen (1986), and the
cooling time at the core radius is less than the Hubble time, so that
the isothermal $\beta$-model is not physically consistent. When we fit
the data with the most recent versions of the cooling-flow models used
by Hardcastle \etal\ (1998d), we find that the dependence of pressure
on radius must be weaker than the $p \propto r^{-1}$ relation used in
our earlier work, because the HRI emission is not strongly centrally
peaked; for example, acceptable fits are obtained for a model with $p
\propto r^{-0.5}$, $\rho \propto r^{-0.75}$, $T \propto r^{0.25}$,
with the cooling time at the core radius being $\sim 2 \times 10^{9}$
yr. This fit is not as good as the simple $\beta$ model together with
a weak point source, and the cooling time is rather short, but it fits
the data without the need for a central radio-related
contribution. For consistency with our other analyses, we use the
point-source component derived from $\beta$-model fitting as our best
estimate of the unresolved contribution to the X-ray emission.

\subsection{3C\,303}

This broad-line radio galaxy is a strong point source in the X-ray. In
the HRI image a faint point-like component, containing $25 \pm 6$
counts, is approximately coincident (after aligning the radio core and
X-ray centroid) with the western hotspot of 3C\,303. We may therefore
be seeing an X-ray counterpart to the radio and optical hotspots
already known in this object (e.g.\ Meisenheimer, Yates \& R\"oser
1997). If so, this would be one of only a very few X-ray hotspots
known. [The other examples are the hotspots of Cygnus A (Harris \etal\
1994) where the emission mechanism is likely to be synchrotron
self-Compton scattering, the western hotspot of Pictor A (R\"oser \&
Meisenheimer 1987, and unpublished {\it ROSAT} data) where the
emission mechanism is not clear, and the northern hot spot of
3C\,390.3 (Prieto 1997, and see below) which is interpreted as a
synchrotron X-ray source.]  However, the position of this X-ray component,
given the low count rate, is at least equally consistent with that of
a $z=1.57$ quasar 4 arcsec to the SW of the optical source identified
with the hot spot (Kronberg 1976; Kronberg \etal\ 1977), and it seems
most likely that the X-ray emission comes from this quasar (see Fig.\
\ref{303-overlay}). If the X-ray emission {\it is} related to the hot
spot, then its 1-keV flux of 7.6 nJy means that it is much too bright
[using the code of Hardcastle, Birkinshaw \& Worrall (1998a)] to be
synchrotron self-Compton emission unless the source is very far from
equipartition. It could be synchrotron emission if the electron energy
distribution is a steep power law (electron energy index of $\sim
2.8$, corresponding to the observed radio and radio-optical spectral
indices of $\sim 0.9$) and extends to very high energies, with a
high-energy cutoff near the region corresponding to X-ray emission.

\subsection{3C\,309.1}

In spite of the suggestion by Forbes \etal\ (1990), based on the
presence of a luminous extended emission-line region, that this quasar
lies in a massive cooling flow, its X-ray emission is well fitted by a
PSPC PRF. As for 3C\,254, we cannot rule out the possibility of a
cooling-flow contribution to the point-source flux, which would need
to be present on smaller scales than the PSPC resolution (i.e. about
100 kpc at the redshift of 3C\,309.1).

\subsection{3C\,310}

This relaxed double radio galaxy (van Breugel \& Fomalont 1984) shows
clear extended X-ray emission, first detected with {\it Einstein} by
Burns, Gregory \& Holman (1981). The HRI image shows the X-ray
emission to be centred on the radio galaxy, with some evidence for
east-west extension in the very central regions; because some other
sources in the field seem to be extended in a similar direction, we
are inclined to treat this as an aspect-induced artefact. When we
characterise the emission with a $\beta$ model and point source (Table
\ref{profile}), $\beta$ is not well constrained but large core radii
(between 40 arcsec for $\beta = 0.35$ and 140 arcsec for $\beta=0.9$)
provide the best fit to the data. No model without a point-source
contribution is a good fit.

\subsection{3C\,324}

We find a very weak extended source close to the catalogued position
in the HRI observation of this high-redshift radio galaxy. Our count
rates for this source are consistent with those of Crawford \& Fabian
(1996a) and Dickinson \etal\ (1998).

\subsection{NGC 6109 (4C\,35.40)}

Our results for this FRI radio galaxy are consistent with those of
Feretti \etal\ (1995), who find a compact source. Radial-profile
fitting suggests that some of the emission is extended (Table \ref{profile}).

\subsection{3C\,334}

This quasar appears point-like in a PSPC observation. A long HRI
observation shows some extension in a NW-SE direction but, as this is
in the same direction as a `hot line' at the edge of the image, it
seems plausible that much of the extension is due to aspect
smearing. However, when the source is dewobbled, radial profiling
reveals some residual extension on scales of tens of arcseconds. The
best-fitting $\beta$ model (Table \ref{profile}) has significant
extended emission. Such a model is consistent with the PSPC radial
profile, although the total count rate in the HRI observation is
considerably higher than would be inferred from the PSPC data,
suggesting variability. There is no evidence in the galaxy counts of
Yee \& Green (1987) that 3C\,334 lies in a particularly rich
environment, but Crawford \& Fabian (1989) show it to have some
extended optical line emission and Hintzen (1984) presents an optical
plate with a number of nearby galaxies, including an apparent close
companion. The compact nature of the extended emission means that
a cooling flow is required unless the temperature is very high.

\subsection{3C\,338}

This radio galaxy, located in the cluster A2199, shows a complex and
asymmetrical X-ray structure (Owen \& Eilek 1998) and is also unusual in
the radio. Radial profiling is clearly not particularly
appropriate, but a simple application to the inner regions of the
source gives the decomposition tabulated here, which represents our
best estimate of the core contribution.

\subsection{NGC 6251}

This FRI radio galaxy shows both extended and compact emission in the
PSPC observation (Birkinshaw \& Worrall 1993; Worrall \& Birkinshaw
1994). The HRI observation detects mainly the compact core, but there
is also a weak but significant detection of a component $\sim 25$
arcsec to the NW of the core, coincident with a bright knot in the
base of the jet (Perley, Bridle \& Willis 1984). [This component is
not to be confused with the possible jet-related X-ray emission claimed
by Mack, Kerp \& Klein (1997), which is at much larger distances from
the nucleus and is not visible on the HRI image.] The knot contributes
only $10 \pm 5$ counts to the total. The number of counts in the
unresolved component is consistent with the results of the spatial
decomposition of Worrall \& Birkinshaw (1994), a result insensitive to
the spectrum of the source. We cannot rule out a thermal contribution
to the unresolved component, but it must be compact.

\subsection{3C\,346}

This narrow-line radio galaxy is clearly detected with the PSPC,
showing both extended and compact X-ray emission; fitting a PSPC PRF
alone to the radial profile gives unacceptable results. In the radio,
3C\,346 is another `jetted double' source with a strong one-sided jet
and a weak hot spot (compare 3C\,288, section \ref{jetted}). It has an
unusually bright core (Spencer \etal\ 1991) and an optical jet (e.g.\
de~Koff \etal\ 1996) and exhibits a strong Laing-Garrington
effect\footnote{Laing (1988); Garrington \etal\ (1988).} (Akujor \&
Garrington 1995) which suggests a cluster environment for the source.
$\beta$-model fits are tabulated in Table \ref{profile}. Most of the
counts (70 per cent) are contributed by the point-source component; no
model without a point-source contribution can adequately fit the
data. A cooling flow is not required.

\subsection{3C\,345}

This superluminal quasar is known to be variable in both the radio and
X-ray (e.g.\ Unwin \etal\ 1997). We use a single PSPC observation,
taken between 1993 Mar 06 and 1993 Mar 11, as representative (since
the available HRI observations are all short). The data are acceptably
fit with a point-source model.

\subsection{3C\,356}

This high-redshift radio galaxy was observed with the {\it ROSAT} PSPC
and HRI by Crawford \& Fabian (1993, 1996b). We confirm the existence
of a very weak source in the HRI observation, and our measurements are
consistent with theirs within the errors.

\subsection{4C16.49}

This steep-spectrum quasar is only barely detected. It appears elongated in
the HRI observation, but in the same direction as other objects in the
field, and so is likely to be aspect-smeared. We treat it as unresolved.

\subsection{3C\,368}

Our results for this high-redshift radio galaxy are consistent with
those of Crawford \& Fabian (1995).

\subsection{3C\,388}

The X-ray images of this low-excitation FRII radio galaxy show a
significant amount of diffuse emission. The source is known to lie in
a poor cluster, with a number of nearby galaxies (Prestage \& Peacock
1988). In the radio the object has an unusual structure. Its one-sided
bright jet and diffuse hot spots might put it in the class of jetted
doubles discussed earlier (section \ref{jetted}), but the presence of
two distinct spectral regimes in its lobes suggests that it may be a
restarting source (Roettiger \etal\ 1994). $\beta$-model fits are
tabulated in Table \ref{profile} and show the X-ray emission to be
dominated by an extended component. No model without a point source is
an acceptable fit to the data, however. Gas of the density implied by
the fitted $\beta$ model should be rapidly cooling, with a mass
deposition rate of $\sim 30 M_\odot$ yr$^{-1}$. When we fit with
cooling-flow models, using a temperature $T \sim 3$ keV outside the cooling
radius, the lack of a strong peak in the central regions (as with
3C\,295) means that no model with $p \propto r^{-1}$ is a good fit; we
find acceptable fits for models with a weaker dependence of $p$ upon
$r$, for example with $p \propto r^{-0.5}$, $\rho \propto r^{-0.75}$,
$T \propto r^{0.25}$, $\beta=0.5$, $r_{\rm core} = 13$ arcsec. The
cooling radius for this model is 9 arcsec and the cooling time at this
radius is about $5 \times 10^9$ yr. A point-like nuclear component is
still required, containing $66 \pm 19$ counts.

\subsection{3C\,390.3}

Leighly \etal\ (1997) discuss 90 separate HRI observations of this
broad-line radio galaxy in the context of a variability study; the
source is shown to be highly variable on time-scales of weeks. We adopt
their average value as our count rate. As discussed by
Harris, Leighly \& Leahy (1998a), the radial profile of the stacked HRI
observations is broader than would be expected from the nominal HRI
PRF, but its width is within the range observed to be a result of
aspect uncertainty, and so we treat the source as point-like [with the
exception of the X-ray hot spot discussed by Prieto (1997) and Harris
\etal ].

\subsection{3C\,442A}

This peculiar radio galaxy (Comins \& Owen 1991) is associated with
NGC 7237 in the merging galaxy pair NGC 7236/7237, which lies in a
poor cluster. The small-scale structure in the HRI image seems to
suggest a double X-ray core with separation roughly 20 arcsec in an
east-west direction. Such structure is unreliable because of aspect
problems, but the large extension makes an aspect-related explanation
unlikely. Other weak sources nearby are not doubled up in the same
way, including one apparently coincident with the small associated
galaxy to the SE. No compact source is found associated with NGC
7236. On larger scales, the map smoothed with a Gaussian of
$\sigma=16$ arcsec seems to show a bar of X-ray emission, oriented
roughly NNW-SSE (see Fig.\ \ref{442a-overlay}); this lies in
approximately the same position angle as the common optical envelope
of the galaxy pair (e.g.\ Borne \& Hoessel 1988) and approximately
perpendicular to the position angle of the radio source, contrary to
the prediction of Comins \& Owen (1991) who expected the hot gas to be
aligned with the radio source. An X-ray counterpart to one of the
nearby radio point sources (`E' in the notation of Comins \& Owen) is
also detected, but source `A' is not seen.

Because of the clear absence of radial symmetry in this source, radial
profiling is not suitable. An upper limit is found for the point-like
counts by taking a 10-arcsec source circle about the X-ray component
coincident with the steep-spectrum radio core, using a background
region between 40 and 60 arcsec away; we correct the result to
our standard 1-arcmin source region. Total counts are derived from a
2.5-arcmin source circle.

\subsection{3C\,449}

The PSPC observations of this FRI radio galaxy are reported in
Hardcastle \etal\ (1998c). The structure of the extended emission is
poorly constrained by a more recent 19-ks HRI observation, but no
point sources are detected which might have conspired to produce the
ring-like structure seen around the southern radio lobe in the lower
resolution PSPC observations, reinforcing the conclusion that this is
caused by a displacement of the hot cluster gas by the radio-emitting
plasma. The HRI and PSPC radial profile fits are consistent, and so we
tabulate the PSPC results in Table \ref{profile}, as the statistics
are slightly better.

\subsection{3C\,454.3}

This is a well-known highly variable core-dominated quasar. We report
results from PSPC observations since the HRI exposures are all
short. The count rates in a 2-arcmin source circle are significantly
different in three separate PSPC observations, and so we have used the longest
(from 1992 Dec 16 -- 1992 Dec 18), which gives a count rate consistent
with that of Sambruna (1997). We find no evidence for intraday
variability or extended X-ray emission.

\subsection{3C\,465}

Sakelliou \& Merrifield (1998) present HRI data which show a
point-like X-ray component coincident with the host galaxy of this
wide-angle tail radio galaxy. After excluding X-ray emission from the
northern companion galaxy of the AGN host and realigning the HRI data
from the two observation epochs as suggested by Sakelliou \&
Merrifield, we find the radial profile of the central source is
moderately well fit by a broadened PRF; we take a 1-arcmin source
circle and background between 1.7 and 4 arcmin, excluding confusing
sources. Further out, there is clear evidence for extended emission on
arcminute scales.

\section{Results}

\subsection{Demographics}

Table \ref{detections} shows the numbers of 3CRR sources observed and
detected in {\it ROSAT} pointed observations, broken down by source
class. Just over half of the 3CRR sample has been observed. FRII radio
galaxies, which make up 60 per cent of the sample, have only 46 per
cent of the observations but are consistent with the overall detection
fraction of 79 per cent in giving 37 per cent of the detections.

The observed subsample is biased with respect to the original (178-MHz
flux density) selection criterion of 3CRR -- the sources that have
been observed with {\it ROSAT} tend to be brighter than those that
have not, largely because the sources with 178-MHz fluxes much larger
than the flux limit of the sample tend to be low-redshift objects and
so more suitable for observation with {\it ROSAT}. However, dividing
the observed subsample by source class, we find that the bias is
only present in the FRII radio galaxies. The subsamples of FRI radio
galaxies and of quasars are unbiased, in the sense of having 178-MHz
radio fluxes that are consistent with having been drawn at
random from the objects of their class in the 3CRR sample. We can
treat at least these subsamples as representative of the 3CRR sample
as a whole.

\subsection{The radio core -- X-ray core correlation}
\label{correlation}

In Table \ref{results} we list for each source the total counts found
in the source region and the number (or upper limit) that we associate with an
unresolved central source. The point-like count rate is
converted into a 1-keV flux density and rest-frame spectral luminosity
density\footnote{The X-ray luminosity density of a
typical galaxy due to intragalactic gas, LMXB and stars, which will be
unresolved in most of our objects, corresponds to only a few $\times
10^{15}$ W Hz$^{-1}$ sr$^{-1}$ at 1 keV (see e.g. Irwin \& Sarazin
1998), and so most of the luminosity of almost all our sources must
arise elsewhere; similarly, all our sources (except M84) have nuclear
luminosities significantly higher than those of the compact nuclear
sources in nearby normal galaxies studied by Colbert \& Mushotzky
(1998).}  by assuming a power-law spectrum (with $\alpha_X =0.8$) and
galactic absorption. We also tabulate 5-GHz radio core flux density
and rest-frame spectral luminosity density, assuming $\alpha_R = 0$. For
consistency we have used the arcsecond-scale rather than
milliarcsecond-scale core fluxes even where both are available; in the
objects (largely quasars) where good information is available the
milliarcsecond- to arcsecond-scale core flux density ratio is not very
different from unity.  Fig.\ \ref{alpha} shows histograms of the
rest-frame radio to core-X-ray two-point spectral indices of sources with
detections in the radio.

A strong correlation is apparent, in both flux-flux and
luminosity-luminosity plots (Figs \ref{ff} and \ref{ll}) between the
X-ray and radio core strengths.  Analysis with the survival analysis
package {\sc asurv} [Rev.~1.1; LaValley, Isobe \& Feigelson (1992)]
taking account of the censoring of the data, using a modified
Kendall's $\tau$ algorithm, shows that the flux-flux and
luminosity-luminosity correlations are significant, both for the
sample as a whole and for the different classes of source (Tables
\ref{fcor} and \ref{lcor}, columns 3 and 4). The fact that the
flux-flux correlation is significant gives us confidence that we are
not simply seeing a redshift-induced artefact in the stronger
luminosity-luminosity correlation. Akritas \& Siebert (1995) present a
procedure which allows partial correlation analysis in the presence of
censoring, and we apply it to the distributions of radio luminosity,
X-ray luminosity and redshift to make this conclusion quantitative; the
correlation for the sample as a whole, and for the radio galaxies and
quasars separately, is significant at better than the 95 per cent
confidence level even given the effects of redshift.

These results imply a physical relationship between the X-ray and
radio cores of the radio sources, confirming work done with other
samples and instruments on low-luminosity radio galaxies (Fabbiano
\etal\ 1984; Edge \& R\"ottgering 1995; Siebert \etal\ 1996; Canosa
\etal\ 1999), high-luminosity radio galaxies (Worrall \etal\ 1994;
Hardcastle \etal\ 1998d) and quasars (Tananbaum \etal\ 1983; Kembhavi,
Feigelson \& Singh 1986; Worrall \etal\ 1987; Browne \& Murphy 1987;
Baker, Hunstead \& Brinkmann 1995; Siebert \etal\ 1996). What does
this imply for the origin of the X-ray emission? 

There is much evidence for relativistic motion in the cores of all
radio-loud AGN, with Lorentz factors $\gamma \ga 5$ inferred from
superluminal motion and unification arguments. The observed fluxes and
luminosities of radio cores are therefore strongly influenced by
Doppler beaming. For example, 20 of the 22 FRI radio galaxies in our
sample, which have broadly similar total radio luminosities, have a
distribution of core prominences (defined in this case as the ratio of
5-GHz core flux density to 178-MHz total flux density) spanning two to
three orders of magnitude and consistent with having been drawn from a
randomly oriented population of objects with a single intrinsic core
prominence and $\gamma = 5$. (The outliers are the peculiar objects
3C\,28 and 3C\,442A, where nuclear activity may have ceased). If the
X-ray luminosity ($L_X$) were simply related to the AGN power, and
therefore independent of beaming and orientation, we would not expect
the core radio and X-ray emission to be strongly correlated. At
best there would be two to three orders of magnitude of scatter in the
correlation, and in practice for 20 objects of similar unbeamed
luminosity we would simply not see the correlation at all. The very
existence of a correlation between $L_X$ and $L_R$ in these radio
galaxies forces us to the conclusion that the soft-X-ray emission
originates in a Doppler-boosted region with a Lorentz factor similar
to that of the jet. The simplest model is one in which the X-ray
emission originates in the jet itself.

To make this argument more quantitative, and to extend it to the
higher-luminosity radio galaxies where there are fewer clear
detections of nuclear X-ray emission, it is useful to characterize the
X-ray/radio correlation by linear regression. Linear regression is
difficult in this dataset and its subsets because not all objects have
detected cores either in the radio or the X-ray wavebands. Most
available algorithms for linear regression in the presence of censored
data work only with unidirectional censoring.  Two well-known
algorithms exist for performing linear regression on doubly censored
data: that of Schmitt (1985), implemented in {\sc asurv}, and the
Theil-Sen estimator (Akritas, Murphy \& LaValley 1995). Their
application to datasets such as ours is discussed in Appendix A. Our
conclusion is that the Theil-Sen estimator is slightly better, but we
tabulate the results of both in Tables \ref{fcor} and \ref{lcor} to
emphasise that similar results are obtained. In a few of the
sub-samples the censoring is only unidirectional (e.g. in quasars
where we have complete radio core information) and we are able to use
other algorithms, such as the Buckley-James method implemented in {\sc
asurv}, which also give similar answers.  The results are most
doubtful in subsamples where upper limits dominate (e.g.\ the FRII
radio galaxies).

Table \ref{lcor} shows that for the subsamples of radio galaxies the
slopes of the correlation are close to unity. Only when broad- and
narrow-line radio galaxies are considered together is the slope
clearly different from unity, and (as can be seen from Fig.\ \ref{ll})
this is because the broad-line objects are systematically brighter in
X-rays than the narrow-line objects of similar radio core luminosity,
suggesting that these objects have an additional X-ray component not
seen in the narrow-line and low-excitation objects. The quasars have
an X-ray-radio slope significantly flatter than unity, and generally
lie about the regression line for the whole sample. This result
has been seen before and can be explained in terms of a model in
which both beamed radio-related X-ray emission and anisotropic
unrelated X-ray emission are seen from quasar nuclei (Worrall \etal\
1987; Browne \& Murphy 1987; Worrall \etal\ 1994; Baker \etal\ 1995).

Of particular interest are the high-power, narrow-line FRII radio
galaxies, of which Cygnus A is the prototype. Obscuring tori are
required by unified models in these sources. Because they are distant
and faint, we have detected only 17 out of the 30 observed sources in
this class, and convincingly detected a central unresolved component
in only 7. However, the survival-analysis slope of the $L_X/L_R$
correlations in these FRIIs (Table \ref{lcor}) is entirely consistent
with the results for the much better detected FRI radio galaxies. If
we normalize the $L_X/L_R$ relation at a 5-GHz radio spectral luminosity
density of $10^{23}$ W Hz$^{-1}$ sr$^{-1}$, then the expected X-ray
luminosity density at 1 keV (with 90 per cent confidence range derived
from simulation) is $1.8 \times 10^{16}$ ($6.5 \times 10^{15}$ -- $4.6
\times 10^{16}$) W Hz$^{-1}$ sr$^{-1}$ for the FRI radio galaxies and
$9.5 \times 10^{15}$ ($1.4 \times 10^{15}$ -- $4.6 \times 10^{16}$) W
Hz$^{-1}$ sr$^{-1}$ for the FRIIs, so the normalizations of the
$L_X/L_R$ relations for the two classes of source are consistent. This
is as expected if the nuclear soft X-ray emission originates from the
jet in both classes of source, since there is little difference on
parsec scales between the radio properties of FRI and FRII jets
(e.g. Pearson 1996). There are no objects which have a bright radio
core and a weak or undetected nuclear X-ray component; therefore
nothing explicitly contradicts the suggestion that there is a
radio-related soft X-ray component in the centre of {\it all} radio
galaxies.

The slope of the radio-X-ray correlations can in principle provide
information on the mechanism responsible for the radio-related X-ray
emission.  Possible mechanisms include synchrotron emission from the
same population of electrons responsible for the radio emission, and
inverse-Compton scattering by those electrons of a population of
low-energy photons to X-ray energies, either the radio synchrotron
photons themselves (synchrotron self-Compton emission) or an external
photon population (`external Compton').  These different mechanisms
would give rise to different relationships between the intrinsic radio
and X-ray luminosities. However, simulation shows that the effects of
Doppler beaming will tend to obscure any relationship that there may
be between the {\it rest-frame} luminosities in the two wavebands. In
addition, redshift-induced effects will tend to push the slope towards
unity. It is conventional to compensate for these problems by considering not
the luminosity but the prominence of a beamed component, where the
prominence is defined as the ratio of the component's luminosity to an
unbeamed quantity such as the low-frequency radio luminosity of the
source. The slope of a plot of X-ray core prominence against radio
core prominence, determined using the Theil-Sen regression, is
consistent with unity for the FRI radio galaxies and the narrow-line
FRII radio galaxies. This suggests that the bulk Lorentz factors of
the radio- and X-ray emitting material are similar.

\subsection{Cluster emission in high-power radio galaxies and quasars}

We have detected extended emission, presumably hot gas associated with
a cluster of galaxies, in several quasars with redshifts between $\sim
0.4$ and $\sim 0.7$ (3C\,48, 3C\,215, 3C\,254, 3C\,275.1, 3C\,334). In
all these objects optical observations have previously suggested a
cluster environment for the radio source. Using the low-redshift
cluster temperature-luminosity relation of David \etal\ (1993), which
where tested has been reliable out to high redshift (Mushotzky \&
Scharf 1997; Donahue \etal\ 1998), we use PROS and the observed count
rates to estimate\footnote{The procedure used is to guess a
temperature, calculate a luminosity from this temperature using a
redshifted Raymond-Smith model with 0.5 cosmic abundance, use the
temperature-luminosity relation to calculate a new temperature from
this luminosity and iterate until the temperature and luminosity
stabilise.} self-consistent temperatures and 2-10 keV luminosities
(see Table \ref{exlumin}). These luminosities are consistent with
moderately rich cluster environments for the quasars, and also match
those of radio galaxies with unambiguously extended X-ray emission at
similar redshift inferred by Worrall \etal\ (1994), Crawford \& Fabian
(1996a), and Hardcastle \etal\ (1998d), as would be expected in
unified models.

Deriving useful upper limits on extended luminosity for those objects
(mostly high-redshift quasars) which appear pointlike in radial
profile analysis is difficult; the answer depends on the radial
profile and the temperature of the undetected extended X-ray emission
as well as on the source and the background count rates. If we assume
that the undetected emission is similar to that in detected objects,
the task is better constrained. X-ray emission from the detected
$z>0.3$ clusters in our sample can be represented on average by a
$\beta$ model with $\beta=0.9$, $r_{\rm core} \sim 150$ kpc, and $kT
\sim 5$ keV. We then use Monte Carlo simulations to estimate how
strong such a component would have to be to be detected at the 95 per
cent significance level on an F-test in 95 per cent of the trials. For
each of the 21 point-like objects with $z>0.3$ we determine an upper
limit on extended counts in this way using a simulation matched to the
on-source and background counts of the source. Where the simulations
show that all the observed counts could be attributed to cluster
emission, as happens in a few cases with poor statistics at high
redshift, we use the total counts as an upper limit instead. From
these limits on count rate we compute a limit on 2-10 keV extended
luminosity assuming $kT = 5$ keV (the results are insensitive to
choice of temperature). Limits are plotted with detections in
Fig.\ \ref{extz}. The upper limits on cluster luminosity are a substantial
fraction of the total source luminosities for the majority of these
objects, since most were observed with the PSPC which has poor spatial
resolution compared with the expected angular size of
clusters. Nevertheless, limits and detections populate similar ranges
of luminosity for high-redshift radio galaxies and quasars.

At lower redshifts, it is known that FRI radio galaxies tend to lie in
group- or cluster-scale hot gas (e.g. Worrall \& Birkinshaw 1994)
and results here agree with this. We cannot rule out extended X-ray
emission in any of the FRIs in the sample, although objects observed
only with the HRI are poorly constrained. The picture is less clear
for FRII objects. Of the low-excitation objects, all seem to lie in
environments that are at least comparable to those of the FRIs, and
some (e.g. 3C\,388) are in considerably richer environments. Of the
high-excitation objects, which are more typical of their class in
their radio structure, at least one (3C\,219) appears to lie in an
environment comparable to 3C\,388's, and some (e.g.\ 3C\,223, 3C\,284)
probably have luminosities comparable to those of the cluster around
the FRI 3C\,449, but others (e.g.\ 3C\,326, 4C\,73.08, 3C\,98) are
constrained by the data to lie in environments no more luminous than
those of the least luminous FRIs. It seems, therefore, that FRIIs at
low redshift inhabit a range of environments, and while we cannot rule
out the possibility that the environments of low-$z$ 3CRR FRIIs are
identical to those of FRIs (although there is optical evidence against
this; Prestage \& Peacock, 1988), our finding that luminous X-ray
environments for low-$z$ FRIIs are not common agrees with earlier work
(e.g. Miller \etal\ 1985). Our data provide some weak support for the
common belief (based on optical work; e.g. Yates, Miller \& Peacock
1989) that there is an evolution in the environments of the most
powerful radio sources between $z=0$ and $z \sim 0.5$, in the sense
that they are more likely to lie in rich clusters at high
redshift.

On the basis of {\it Einstein} observations Miller \etal\ (1985)
suggested that the hot-gas environments of low-$z$ FRIIs do not
confine their radio lobes. Our limits on extended emission in the
environments of FRIIs, however, suggest that lobe confinement is
possible; for example, the very poor environment that we detect around
3C\,98, with an assumed temperature of 1 keV, provides a pressure
greater than or equal to the minimum pressure derived for 3C\,98 by
Miller \etal\ out to $\sim 2$ arcmin (i.e. close to the hot spot
regions where ram pressure confinement is believed to be
important). Part of the argument of Miller \etal\ was based on the
observation that sources with a wide range of linear sizes have
approximately the same axial ratio (defined as the ratio of lobe width
to length); this would not happen if the lobes were confined while the
sources continued to expand linearly. But more recent studies of
radio-galaxies at low redshift do show a weak relationship
between source length and axial ratio, removing some of the motivation
for believing that the lobes of radio galaxies are unconfined; it
seems more likely that they are unconfined initially and reach
equilibrium at a later stage in a source's lifetime.

We can compare the clusters detected by X-ray
observation with those inferred from observations of extended
emission-line gas. Emission-line regions imply that there is cool
($10^4$ K) gas in the environment of the quasar or radio galaxy, and
this must be in pressure equilibrium with the hot, X-ray emitting gas
if the emission-line region is to last more than a sound-crossing
time. Pressures can be inferred from emission-line ratios if it is
assumed that the lines are excited by photoionization from the central
quasar. The assumption that the emission-line regions are not {\it
overpressured} with respect to the X-ray emitting gas leads to the
deduction in many objects (e.g. Crawford \& Fabian 1989) that the
X-ray gas must be so dense that its local cooling time is less
than the Hubble time. There are luminous extended emission-line
regions in all the quasars with detected extended X-ray emission, although
the X-ray observations do not directly require cooling in all
cases. Pressures deduced from line diagnostics ([OIII]/[OII] ratios)
for four of these sources are tabulated in Table \ref{exlumin},
together with pressures at similar radii derived from the
$\beta$-model fits to the extended X-ray emission. The pressures are
consistent to within a factor of a few in all cases where data are
available.

After submission of the first draft of this paper, Crawford \etal\
(1999) completed an independent radial-profile analysis of all five of
the quasars in our sample with detected extended emission. They reach
broadly similar conclusions.

\section{Conclusions}

We have used the spatial resolution of {\it ROSAT} to find
compact central X-ray components in a large fraction ($\sim 60$ per
cent) of the observed radio galaxies and quasars in the 3CRR sample,
and we report limits in the remaining sources. It appears that the model,
discussed in section \ref{intro}, in which all powerful radio sources
have radio-related soft X-ray emission is consistent with these
observations. As argued in section \ref{correlation}, the strong correlation between the core radio and nuclear
X-ray emission, given that the radio core is believed
to be strongly affected by relativistic beaming, implies that the
soft X-ray emission originates in the jet, rather than being related
to the radio emission less directly (e.g. by common correlation with
the bolometric luminosity of the AGN). A jet origin for the soft X-ray
emission also explains the fact that it is seen in radio galaxies in
spite of the high column densities of absorbing material inferred from
hard X-ray observations of some sources. A second nuclear X-ray
component, which originates much closer to the nucleus and is affected
by the absorbing material so that it is not seen in radio galaxies, is
necessary to explain the otherwise anomalously bright emission from
lobe-dominated quasars (and from radio-quiet quasars) as discussed by
Worrall \etal\ (1994); such a component is also likely to be present
in broad-line radio galaxies.  The slope of the $L_X/L_R$ correlation
for radio galaxies, qualitatively similar to that seen in
core-dominated quasars, supports unified models for radio galaxies and
quasars.

Extended X-ray emission from cluster- or group-scale gas is common in
FRI objects in the sample. Several low-redshift FRIIs also show
extended X-ray emission, but others must inhabit relatively poor
environments. At higher redshifts, we have found evidence for luminous
X-ray emission around several radio galaxies and
lobe-dominated quasars, implying rich cluster
environments for these sources. The resolution of {\it ROSAT}
limits our ability to detect more quasar clusters, but it is very
likely that more will be detected with the new generation of X-ray
instruments. The inferred environments of radio galaxies and quasars
detected at $z \ga 0.3$ are similar, supporting the predictions of
unified models.

\section*{Acknowledgements}
We thank Mark Birkinshaw for writing and supporting the software used
in radial profile analysis, and for helpful comments on the paper.
Fortran code implementing the Theil-Sen estimator was written by
Michael LaValley and kindly provided by Michael Akritas; we thank him
and Eric Feigelson for helpful discussion of the linear regression
problems discussed in section \ref{correlation} and Appendix
\ref{regression}.  Fortran code implementing the partial Kendall's
$\tau$ procedure of Akritas \& Seibert was also provided by Michael
Akritas, director of the Statistical Consulting Center for Astronomy
operated at the Department of Statistics, Penn State University (URL:
$<$http://www.stat.psu.edu/scca/homepage.html$>$). We are grateful to
Dan Harris for a thorough and wide-ranging referee's report on the
first version of this paper.

This research has made use of data obtained through the High Energy
Astrophysics Science Archive Research Center Online Service, provided
by the NASA/Goddard Space Flight Center. The Digitized Sky Surveys
were produced at the Space Telescope Science Institute under
U.S. Government grant NAG W-2166. The National Radio Astronomy
Observatory is operated by Associated Universities Inc., under
co-operative agreement with the National Science Foundation. This work
was supported by PPARC grant GR/K98582 and NASA grant NAG5-1882.

\appendix
\section{Linear regression in doubly censored data}
\label{regression}

Linear regression when the data are censored in both directions is a
common problem in astronomy, but there are relatively few algorithms
for dealing with it; most of the survival-analysis techniques used in
astronomy are inherited from the biological sciences where only one
variable (survival time) is likely to be censored. Schmitt (1985)
presents one algorithm, which is implemented in the survival-analysis
package {\sc asurv}. Some of the problems with Schmitt's algorithm are
discussed in Sadler, Jenkins \& Kotanyi (1989). One particular
practical difficulty is that the data must be binned, and the correct
procedure for choosing the number of bins is not obvious. Akritas
\etal\ (1995) present a second algorithm, a generalised version of the
Theil-Sen estimator [which determines the slope $\beta$ which makes
Kendall's $\tau$ between the residuals ($y_i - \beta x_i$) and $x_i$
approximately zero]. Here we discuss Monte Carlo simulations intended
to demonstrate which of the two procedures is best suited to our data.

Our aim is to find the unknown physical relationship between two
quantities, both of which probably have intrinsic scatter; this
differs from the more usual case where the error in one of the
quantities (the independent variable) can be neglected. The
measurement errors in our data are probably small compared to the
intrinsic scatter, and are in any case not well known in the case of the
radio core flux densities. In the case of uncensored data of this
kind, where ordinary least-squares regression is applicable, Isobe
\etal\ (1990) have shown that the best estimators of the slope are
those that treat the two variables symmetrically, so that one quantity
is not privileged by being treated as the independent variable; they
recommend the line that bisects the two lines obtained by ordinary
least squares regression of each variable on the other. Schmitt's
algorithm is essentially a weighted least-squares estimator, so it
seems likely that the same problems obtain and that similar solutions
are applicable; the situation is less clear in the case of the
Theil-Sen estimator.

Our simulated data consisted of a hundred data points. The $x_i$ were
drawn from a uniform distribution between 21.0 and 28.0; the
corresponding points $y_i$ were generated as $y_i = \alpha + \beta
x_i$, where $\alpha=-10$. These values were chosen so as to match
roughly with the ranges seen in radio and X-ray luminosity.  Intrinsic
scatter was modelled with Gaussians having standard deviations of
$\sigma_x$ and $\sigma_y$; random numbers drawn from these
distributions were added to $x_i$ and $y_i$. Each point had
independent probabilities $p_x$, $p_y$ of being censored in the $x$
and $y$ directions, which were of the order of tens of per cent to
match our dataset. Censored data were replaced with an upper limit
which was higher by the modulus of a value drawn from a Gaussian
distribution than the value after intrinsic scatter had been added. We
ran a number of simulations (typically several hundred) and measured
the median and 90 per cent confidence levels of the estimates of the
slope. For the Schmitt algorithm the data were binned in a $5 \times
5$ grid with the bin boundaries being chosen automatically by {\sc
asurv}. Simulation showed that the results, and confidence ranges,
were not strongly affected by other choices of number of bins around
this value. Neither estimator was strongly affected by choices of
other distributions for the initial values of $x_i$.

Our modelling shows that the Schmitt procedure results in a biased
estimator of the true slope, $\beta$, of such datasets even if no
scatter is added to the $x$-coordinate, taken to be the independent
variable; it consistently underestimates $\beta$. Only in the limit of
negligible scatter and negligible censoring is it unbiased. By contrast,
the Theil-Sen procedure gives a good, unbiased estimator of $\beta$
when no scatter is added to the $x_i$, as described by Akritas
\etal\ Both, however, are biased (in the sense of underestimating $\beta$)
when intrinsic scatter is added to the dependent variable.

Treating the variables symmetrically by finding the bisector of the
two regression lines has better results. Both the Schmitt and the
Theil-Sen procedures provide essentially unbiased estimators of
$\beta$ provided that there is similar noise and censoring in the two
co-ordinates; `similar' here implies that $\sigma_y = \beta
\sigma_x$. The Theil-Sen procedure deals slightly better with cases
where $\beta \neq 1$, though both tend to produce results that are
biased towards an estimated slope of unity; we emphasise that this
bias is small compared with the breadth of the distribution of the
results in small datasets like ours. The 90 per cent confidence
ranges are slighly smaller when the Theil-Sen estimator is
used. Neither test appears to be biased by allowing the censoring
probability to depend on the value of $x_i$ or $y_i$, so long as this
is done symmetrically. If the censoring probabilities or the degrees
of intrinsic scatter are different in the two co-ordinates then there
is some bias in the results of either algorithm, in the senses
tabulated in Table \ref{bias}. Bias is particularly serious when the
degrees of intrinsic scatter are significantly different; for example,
when $\sigma_x = 2.0$ and $\sigma_y = 1.0$ the median estimate of the
slope is biased by $\sim 20$ per cent, even when censoring is
negligible.

We conclude that
\begin{itemize}
\item The Theil-Sen estimator is undoubtedly better than the
Schmitt estimator when there is negligible intrinsic scatter in one
co-ordinate but non-zero scatter in the other.
\item When the bisector of the two regression lines is used and when
the variables are genuinely symmetrical, both
procedures provide good estimates of the true slope for $\beta \sim
1$. The Theil-Sen estimator gives a smaller confidence range, is less
badly affected by choices of $\beta$ significantly different from 1,
and is less biased by asymmetry in the censoring of the two variables.
\item The bisector of the two regression lines is a significantly
biased estimator of the true slope, using either estimator, when the
noise on the two variables is significantly differently distributed.

\bsp
\clearpage
\renewcommand{\thefigure}{\arabic{figure}}
\renewcommand{\thetable}{\arabic{table}}

\begin{table*}
\caption{3CRR objects in the {\it ROSAT} public archive}
\label{sources}
\begin{tabular}{lrrrrlrr}
Source&$z$&FR class&Source type&Galactic $N_H$&ROR
number&Livetime&Offset\\
&&&&($\times 10^{20}$ cm$^{-2}$)&&(s)&(arcmin)\\
\hline
3C\,13&1.351 &II&NLRG&6.24 &rp600244n00 &34680 &29 \\
3C\,20&0.174 &II&NLRG&17.93 &rh702084n00 &5138 &-- \\
3C\,28&0.1952 &I&LERG&5.14 &rh800633 &50266 &-- \\
3C\,31&0.0167 &I&LERG&5.53 &rh600496 &24815 &-- \\
3C\,33&0.0595 &II&NLRG&3.21 &rh7020872 &47685 &-- \\
3C\,33.1&0.181 &II&BLRG&22.50 &rh702063 &18263 &-- \\
3C\,47&0.425 &II&Q&5.34 &rp700069, rp700853 &10835 &-- \\
3C\,48&0.367 &C(CSS)&Q&4.50 &rh800634 &37020 &-- \\
3C\,61.1&0.186 &II&NLRG&7.57 &rh702064n00 &15943 &-- \\
3C\,66B&0.0215 &I&LERG&6.84 &rh700365n00 &10207 &-- \\
3C\,67&0.3102 &II&BLRG&7.12 &rp300324n00 &2782 &24 \\
3C\,79&0.2559 &II&NLRG&10.09 &rh701607n00 &881 &-- \\
3C\,83.1B&0.0255 &I&LERG&13.86 &rh701714n00 &37680 &-- \\
3C\,84&0.0172 &I&NLRG&13.83 &rh800591a01 &51913 &-- \\
3C\,98&0.0306 &II&NLRG&11.70 &rh701614a02 &41047 &-- \\
3C\,123&0.2177 &II&LERG&18.29 &rh704034n00 &28801 &-- \\
3C\,171&0.2384 &II&NLRG&6.60 &rh702623n00 &20690 &-- \\
3C\,181&1.382 &II&Q&7.59 &rh701944n00 &1594 &-- \\
3C\,192&0.0598 &II&NLRG&5.06 &rh700367m00 &15954 &-- \\
3C\,196&0.871 &II&Q&4.93 &rp700249n00 &6294 &-- \\
3C\,204&1.112 &II&Q&4.85 &rp201472n00 &4811 &15.3 \\
3C\,207&0.684 &II&Q&5.40 &rp701170n00 &6763 &-- \\
3C\,208&1.109 &II&Q&3.60 &rp700887n00 &18068 &-- \\
3C\,212&1.049 &II&Q&4.09 &rp700436n00 &20888 &-- \\
3C\,215&0.411 &II&Q&3.75 &rh800718, rh800753 &86442 &-- \\
3C\,216&0.668 &C(CSS)&Q&1.40 &rp700329a01 &21346 &-- \\
3C\,219&0.1744 &II&BLRG&1.48 &rp700539n00 &4206 &-- \\
3C\,220.1&0.61 &II&NLRG&1.93 &rh701727n00 &36226 &-- \\
3C\,220.3&0.685 &II&NLRG&2.40 &rp700072n00 &8791 &-- \\
3C\,223&0.1368 &II&NLRG&1.20 &rp700389n00 &7817 &-- \\
4C\,73.08&0.0581 &II&NLRG&2.45 &rp701214n00 &12010 &-- \\
3C\,236&0.0989 &II&LERG&1.26 &rp701215 &11946 &-- \\
3C\,241&1.617 &II&NLRG&2.02 &rp700050 &8293 &-- \\
3C\,245&1.029 &II&Q&2.70 &rp700384n00 &9735 &-- \\
3C\,247&0.7489 &II&NLRG&1.03 &rh702715n00 &31137 &-- \\
3C\,249.1&0.311 &II&Q&2.89 &rp700070n00 &1662 &-- \\
3C\,254&0.734 &II&Q&1.75 &rp700855n00 &15570 &-- \\
3C\,263&0.6563 &II&Q&0.91 &rh800027a01 &31203 &-- \\
3C\,263.1&0.824 &II&NLRG&2.09 &rh702716a01 &33457 &-- \\
3C\,264&0.0208 &I&LERG&2.20 &rp800153n00 &18141 &8.4 \\
3C\,266&1.2750 &II&NLRG&1.69 &rp201237 &5767 &17 \\
3C\,268.3&0.371 &II&BLRG&1.88 &rh800442n00 &6258 &-- \\
3C\,268.4&1.400 &II&Q&1.25 &rh701943n00 &5718 &-- \\
3C\,270.1&1.519 &II&Q&1.14 &rp700864 &21584 &-- \\
3C\,272.1&0.0031 &I&LERG&2.78 &rh600493n00 &26237 &-- \\
A1552&0.0837 &I&LERG&2.43 &rp800577n00 &3259 &-- \\
3C\,274&0.0043 &I&NLRG&2.51 &rh701712n00 &44264 &-- \\
3C\,275.1&0.557 &II&Q&1.89 &rh800719n00 &25158 &-- \\
3C\,277.2&0.766 &II&NLRG&1.93 &rp800393a01 &13141 &-- \\
3C\,280&0.996 &II&NLRG&1.25 &rp700073n00 &46619 &-- \\
3C\,284&0.2394 &II&NLRG&1.0 &rp201471n00 &8374 &27 \\
3C\,288&0.246 &I&LERG&0.81 &rp900222 &7793 &21 \\
3C\,289&0.9674 &II&NLRG&1.21 &rh702713n00 &29736 &-- \\
3C\,293&0.0452 &I&LERG&1.29 &rh700366 &10104 &-- \\
3C\,294&1.78 &II&NLRG&1.20 &rh800806n00 &93123 &-- \\
3C\,295&0.4614 &II&NLRG&1.38 &rh800678 &29292 &-- \\
3C\,296&0.0237 &I&LERG&1.85 &rh701833n00 &29162 &-- \\
3C\,299&0.367 &II&NLRG&1.20 &rh702714n00 &59927 &-- \\
3C\,303&0.141 &II&BLRG&1.60 &rh701723 &22075 &-- \\
3C\,309.1&0.904 &C(CSS)&Q&2.41 &rp700330n00 &9472 &-- \\
3C\,310&0.0540 &I&LERG&3.42 &rh702069 &27756 &-- \\
\end{tabular}
\end{table*}
\begin{table*}
\contcaption{}
\begin{tabular}{lrrrrlrr}
Source&$z$&FR class&Source type&Galactic $N_H$&ROR
number&Livetime&Offset\\
&&&&($\times 10^{20}$ cm$^{-2}$)&&(s)&(arcmin)\\
\hline
3C\,318&0.752 &II&LERG&3.90 &rp201179n00 &1195 &20 \\
3C\,324&1.2063 &II&NLRG&4.47 &rh701720n00 &71603 &-- \\
3C\,326&0.0895 &II&NLRG&3.87 &rp701213 &14249 &-- \\
3C\,325&0.86 &II&Q&1.74 &rh701605n00 &28168 &-- \\
3C\,330&0.5490 &II&NLRG&2.94 &rp800149 &3978 &14 \\
NGC\,6109&0.0296 &I&LERG&1.47 &rh800164n00 &33399 &-- \\
3C\,334&0.555 &II&Q&4.14 &rh800720n00 &27909 &-- \\
3C\,338&0.0298 &I&NLRG&0.88 &rh800429a01 &20755 &-- \\
3C\,343&0.988 &C(CDQ)&Q&2.45 &rp701171n00 &4653 &-- \\
3C\,343.1&0.750 &II&NLRG&2.68 &rp701171n00 &4653 &29 \\
NGC\,6251&0.024 &I&LERG&5.82 &rh701306n00 &17387 &-- \\
3C\,346&0.162 &I&NLRG&5.47 &rp701409n00 &16981 &-- \\
3C\,345&0.594 &C(CDQ)&Q&0.90 &rp700870n00 &5703 &-- \\
3C\,351&0.371 &II&Q&2.03 &rp701439n00 &14804 &-- \\
3C\,356&1.079 &II&NLRG&2.74 &rh800717n00 &36028 &-- \\
4C\,16.49&1.296 &II&Q&6.54 &rh701945n00 &9939 &-- \\
3C\,368&1.132 &II&NLRG&9.80 &rp880394n00 &25429 &-- \\
3C\,380&0.691 &C(CSS)&Q&6.60 &rp700250n00 &6209 &-- \\
3C\,382&0.0578 &II&BLRG&6.70 &rh700368 &4943 &-- \\
3C\,388&0.0908 &II&LERG&5.81 &rh701613n00, rh701722n00 &52674 &-- \\
3C\,390.3&0.0569 &II&BLRG&3.74 &many HRI &1900 &-- \\
3C\,433&0.1016 &II&NLRG&9.15 &rp700519n00 &792 &-- \\
3C\,442A&0.0263 &I&LERG&5.08 &rh701835n00 &9060 &-- \\
3C\,449&0.0171 &I&LERG&11.05 &rp700886n00 &9151 &-- \\
3C\,454&1.757 &C(CSS)&Q&5.84 &rh701942n00 &2376 &-- \\
3C\,454.3&0.859 &C(CDQ)&Q&7.06 &rp900339n00 &21267 &-- \\
3C\,455&0.5427 &II&Q&4.97 &rh701941n00 &2284 &-- \\
3C\,465&0.0293 &I&LERG&4.91 &rh800715 &61933 &-- \\
\end{tabular}
\vskip 10pt
\begin{minipage}{16cm}
Column 3 lists the source structure classification; most sources are
FRI or FRII (Fanaroff \& Riley 1974). A C in this column indicates
compact or core-dominated structure; such objects are subdivided into
compact steep-spectrum (CSS) or core-dominated, flat-spectrum quasars
(CDQ) according to whether their low-frequency spectral index
($\alpha_{750}^{178}$) is greater or less than 0.5. Letters in column
4 represent optical class; LERG indicates a low-excitation radio galaxy
[by the definition of Laing \etal\ (1994); see the text], NLRG a
narrow-line radio galaxy, BLRG a broad-line radio galaxy and Q a
quasar. The data in these two columns are taken from Laing \& Riley,
in preparation. Galactic neutral hydrogen column densities, in column
5, are taken from Elvis, Lockman \& Wilkes (1989), Danly \etal\
(1992), Lockman \& Savage (1995), and Murphy \etal\ (1996) where
available (mostly for quasars) and otherwise interpolated from Stark
\etal\ (1992). The second character of the ROR number (in column 6)
indicates whether the data used in our analysis were taken with the
HRI or PSPC. The offset in column 8 is the distance between the radio
core position of the target source and the pointing centre of the
observations, where this was significant (i.e. $>1$ arcmin).
\end{minipage}
\end{table*}

\begin{table*}
\caption{Results of radial profile fitting for extended sources}
\label{profile}
\begin{tabular}{lrrrrrrrrrrr}
Source&Luminosity&Source&Background&$\chi^2$&dof&$\beta$&Core&Central surface&Counts
in&Counts in\\
&distance&radius&radiuss&&&&radius&brightness (10$^{-6}$&point
model&$\beta$ model\\
&(Gpc)&(arcsec)&(arcsec)&&&&(arcsec)&counts s$^{-1}$ arcsec$^{-2}$)\\
\hline
3C\,28&1.286 &64 &128 &14.57 &9 &0.67 &15 &33.0 (H)&29 &2040 \\
3C\,48&2.61 &150 &210 &6.42 &13 &0.9 &15 &12.9 (H)&5892 &285 \\
3C\,98&0.186 &180 &240 &8.67 &7 &0.9 &100 &0.1418 (H)&20 &136 \\
3C\,215&2.97 &150 &210 &7.68 &14 &0.9 &20 &2.494 (H)&3915 &301 \\
3C\,219&1.138 &112.5 &150 &2.6 &8 &0.9 &30 &12.8 (P)&360 &129 \\
3C\,220.1&4.8 &60 &90 &0.689 &6 &0.9 &13 &9.72 (H)&117 &152 \\
3C\,254&6.02 &150 &210 &14.3 &16 &0.9 &35 &5.403 (P)&1474 &333 \\
3C\,264&0.126 &120 &180 &21.6 &14 &0.67 &10 &18.5 (P)&3722 &1257 \\
3C\,272.1&0.019 &100 &150 &8.96 &8 &0.9 &20 &48.06 (H)&84 &1301 \\
3C\,275.1&4.27 &150 &210 &1.9 &12 &0.9 &25 &1.96 (H)&367 &81 \\
3C\,280&8.95 &120 &260 &-- &7 &0.67 &65 &0.16 (P)&43 &149 \\
3C\,295&3.407 &150 &210 &16.7 &12 &0.67 &8 &59.4 (H)&23 &675 \\
3C\,310&0.333 &320 &370 &7.17 &9 &0.5 &75 &0.912 (H)&93 &1190 \\
NGC\,6109&0.180 &60 &120 &1.45 &5 &0.9 &25 &1.01 (H)&59 &46 \\
3C\,334&4.25 &60 &120 &4.84 &9 &0.9 &20 &1.689 (H)&1075 &48 \\
3C\,346&1.05 &180 &210 &15.5 &14 &0.67 &60 &1.60 (P)&767 &454 \\
3C\,388&0.570 &210 &240 &16.6 &14 &0.5 &13 &13.3 (H)&99 &2211 \\
3C\,449&0.103 &700 &1000 &7.7 &11 &0.35 &35 &3.09 (P)&33 &1807 \\
\end{tabular}
\vskip 10pt
\begin{minipage}{16cm}
An H in column 9 denotes a surface brightness derived from HRI
observations; a P denotes PSPC results.
\end{minipage}
\end{table*}

\begin{table*}
\caption{Observations and detections of 3CRR sources}
\label{detections}
\begin{tabular}{llrrrr}
Source type&&Observed&Detected&Total in 3CRR&Median 0.2--2.5 keV\\
&&&&&flux (ergs cm$^{-2}$ s$^{-1}$)\\
\hline
FRI radio galaxies&&20& 19& 26$^a$&    1.3$\times 10^{-12}$\\[5pt]
FRII radio galaxies&Low-excitation&4& 3& 16& --\\
&Narrow-line&30& 17& 76&    2.0$\times 10^{-14}$\\
&Broad-line&7& 6& 10&    7.4$\times 10^{-13}$\\
&Total&41& 26& 103$^b$&    3.6$\times 10^{-14}$\\[5pt]
Quasars&FRII&20& 19& 30&    6.4$\times 10^{-13}$\\
&Core-dominated&8& 6& 13&    1.5$\times 10^{-12}$\\
&Total&28& 25& 43&    6.7$\times 10^{-13}$\\[5pt]
All&&89& 70& 172&    3.0$\times 10^{-13}$\\[5pt]
\end{tabular}
\begin{minipage}{14cm}
$^{a}$ The starburst galaxy 3C\,231 (M82) is omitted from the total.\par $^{b}$
One unclassified object (4C\,74.16), not observed with {\it ROSAT}, is
included in the total.\par
Median fluxes tabulated in column 5 take into account non-detections;
they are derived from the Kaplan-Meier
estimator calculated using {\sc asurv}, and are presented to assist
observers in estimating the detectability of such sources.  There are too few
detected low-excitation FRII radio galaxies to allow a meaningful calculation.
\end{minipage}
\end{table*}

\begin{table*}
\caption{Counts and flux densities for X-ray observations and radio
cores}
\label{results}
\begin{tabular}{lrrrrrrrrrr}
&&&&&\multicolumn{2}{c}{Unresolved}\\
Source&\multicolumn{2}{c}{Total}&\multicolumn{2}{c}{Unresolved}&1-keV
flux&$\log_{10}$(1-kev&Approx.&5-GHz core flux&$\log_{10}$(5-GHz&Ref.     \\
&Counts&Error&Counts&Error&density (nJy)&lum.\ density,&unres.&density(mJy)&lum.\ density,\\
&&&&&&W Hz$^{-1}$ sr$^{-1}$)&size (kpc)&&W Hz$^{-1}$ sr$^{-1}$)\\
\hline
3C\,13&$<$45 &-- &$<$45 &-- &$<$4.7 &$<$ 18.84&150 &0.18 $^*$& 23.11&3 \\
3C\,20&$<$6 &-- &$<$6 &-- &$<$20 &$<$ 17.30&12 &2.6 & 22.43&5 \\
3C\,28&7820 &370 &$<$29 &18 &$<$5.4 &$<$ 16.91&13 &$<$0.4 &$<$ 21.70&1 \\
3C\,31&165 &17 &165 &17 &63.7 & 15.79&1.4 &92 & 21.94&1 \\
3C\,33&140 &30 &140 &30 &24.3 & 16.49&4.8 &24 & 22.46&1 \\
3C\,33.1&26 &8 &$<$26 &8 &$<$22 &$<$ 17.46&12 &20.4 $^*$& 23.36&6 \\
3C\,47&1567 &41 &1567 &41 &493.0 & 19.62&96 &73.6 & 24.67&4 \\
3C\,48&5999 &166 &5892 &93 &1440 & 19.94&20 &896 & 25.63&7 \\
3C\,61.1&$<$8 &-- &$<$8 &-- &$<$5 &$<$ 16.85&13 &6.0 $^*$& 22.86&8 \\
3C\,66B&141.6 &33 &141.6 &33 &141.0 & 16.36&1.8 &182 \dag& 22.46&9 \\
3C\,67&22 &9 &$<$22 &9 &$<$30 &$<$ 18.08&79 &5.5 & 23.26&10 \\
3C\,79&$<$2 &-- &$<$2 &-- &$<$30 &$<$ 17.85&16 &10 & 23.36&1 \\
3C\,83.1B&80 &30 &80 &30 &27 & 15.79&2.1 &40 & 21.95&11 \\
3C\,84&321465 &909 &5285 &1057 &1300 & 17.13&1.5 &59600 \dag& 24.78&12 \\
3C\,98&48 &27 &20 &8 &5.8 & 15.28&2.6 &9.0 & 21.46&1 \\
3C\,123&216 &59 &$<$19 &7 &$<$9.4 &$<$ 17.26&14 &100 & 24.21&22 \\
3C\,171&$<$8 &-- &$<$8 &-- &$<$4 &$<$ 16.95&15 &2.2 & 22.63&1 \\
3C\,181&7 &4 &$<$7 &-- &$<$50 &$<$ 19.85&36 &6 & 24.70&2 \\
3C\,192&$<$8 &-- &$<$8 &-- &$<$5 &$<$ 15.78&4.8 &8.0 & 21.99&1 \\
3C\,196&38 &10 &38 &10 &20 & 18.97&140 &7 & 24.30&13 \\
3C\,204&134 &13 &134 &13 &91.6 & 19.91&150 &26.9 & 25.12&4 \\
3C\,207&444 &23 &444 &23 &225 & 19.77&120 &510 \dag& 25.94&2 \\
3C\,208&572 &26 &572 &26 &92.6 & 19.91&150 &51.0 & 25.39&4 \\
3C\,212&749 &31 &749 &31 &110 & 19.92&140 &150 & 25.81&2 \\
3C\,215&3973 &231 &3915 &85 &391.0 & 19.49&22 &16.4 & 23.99&4 \\
3C\,216&1152 &44 &1152 &44 &105.0 & 19.41&120 &1050 & 26.23&2 \\
3C\,219&480 &27 &360 &32 &171 & 18.31&52 &51 & 23.72&1 \\
3C\,220.1&260 &70 &117 &20 &23.5 & 18.67&27 &25 & 24.53&1 \\
3C\,220.3&$<$32 &-- &$<$32 &-- &$<$9.1 &$<$ 18.38&120 &$<$1.0 &$<$ 23.30&26 \\
3C\,223&76 &19 &$<$76 &19 &$<$18 &$<$ 17.11&43 &9 & 22.78&1 \\
4C\,73.08&107 &21 &$<$107 &21 &$<$22.1 &$<$ 16.43&20 &11.2 $^*$& 22.11&14 \\
3C\,236&50 &23 &$<$50 &23 &$<$7.8 &$<$ 16.45&32 &84 & 23.45&1 \\
3C\,241&53 &17 &$<$53 &17 &$<$15 &$<$ 19.54&160 &3 & 24.48&1 \\
3C\,245&626 &34 &626 &34 &166 & 20.08&140 &910 \dag& 26.57&2 \\
3C\,247&$<$10 &-- &$<$10 &-- &$<$2.0 &$<$ 17.82&29 &3.5 & 23.86&1 \\
3C\,249.1&411 &23 &411 &23 &659 & 19.44&79 &71 & 24.38&4 \\
3C\,254&1783 &53 &1474 &66 &203.0 & 19.80&130 &19 & 24.58&2 \\
3C\,263&3070 &83 &3070 &83 &606.0 & 20.15&28 &157 & 25.39&4 \\
3C\,263.1&144 &27 &144 &27 &32.0 & 19.12&31 &3.2 & 23.91&18 \\
3C\,264&4848 &87 &3722 &97 &486.0 & 16.86&7.6 &200 & 22.47&15 \\
3C\,266&$<$11 &-- &$<$11 &-- &$<$4.0 &$<$ 18.71&150 &$<$0.6 &$<$ 23.60&1 \\
3C\,268.3&$<$6 &-- &$<$6 &-- &$<$7 &$<$ 17.60&20 &$<$130 &$<$ 24.80&1 \\
3C\,268.4&69 &11 &69 &11 &79 & 20.11&36 &50 & 25.60&2 \\
3C\,270.1&1010 &42 &1010 &42 &84.40 & 20.23&160 &190 \dag& 26.26&2 \\
3C\,272.1&1288 &68 &84 &21 &26 & 13.92&0.27 &180 & 20.77&1 \\
A1552&$<$9 &-- &$<$9 &-- &$<$7 &$<$ 16.30&28 &27 $^*$& 22.81&25 \\
3C\,274&115338 &875 &6134 &97 &1080 \dag& 15.84&0.37 &4$\times 10^{3}$ & 22.48&1 \\
3C\,275.1&405 &31 &367 &24 &106 & 19.23&26 &130 \dag& 25.16&2 \\
3C\,277.2&11.4 &7.3 &$<$11.4 &7.3 &$<$1.96 &$<$ 17.83&130 &0.36 $^*$& 22.89&3 \\
3C\,280&157 &80 &43 &13 &1.7 & 18.04&140 &1.0 & 23.60&1 \\
3C\,284&50 &16 &$<$50 &16 &$<$10 &$<$ 17.38&66 &3.2 & 22.80&1 \\
3C\,288&750 &60 &$<$750 &60 &$<$154 &$<$ 18.59&67 &30 & 23.80&1 \\
3C\,289&14 &7 &$<$14 &7 &$<$3.1 &$<$ 18.28&32 &$<$0.5 &$<$ 23.30&1 \\
3C\,293&10 &5 &$<$10 &5 &$<$6.6 &$<$ 15.68&3.7 &100 & 22.85&1 \\
3C\,294&47 &13 &$<$47 &13 &$<$3.3 &$<$ 19.00&38 &$<$1.0 &$<$ 24.00&16 \\
3C\,295&693 &77 &23 &20 &5.3 & 17.73&23 &3 $^*$& 23.30&17 \\
3C\,296&234 &30 &234 &30 &57.8 & 16.05&2.0 &77 & 22.18&1 \\
3C\,299&$<$16 &-- &$<$16 &-- &$<$1.7 &$<$ 17.04&20 &1.0 & 22.70&18 \\
3C\,303&1005 &40 &1005 &40 &317.0 & 18.38&10 &150 & 24.01&1 \\
3C\,309.1&984 &36 &984 &36 &256 & 20.12&140 &2350 & 26.86&2 \\
3C\,310&1212 &276 &93 &15 &28 & 16.46&4.3 &80 & 22.90&1 \\
\end{tabular}
\end{table*}
\begin{table*}
\contcaption{}
\begin{tabular}{lrrrrrrrrrr}
&&&&&\multicolumn{2}{c}{Unresolved}\\
Source&\multicolumn{2}{c}{Total}&\multicolumn{2}{c}{Unresolved}&1-keV
flux&$\log_{10}$(1-kev&Approx.&5-GHz core flux&$\log_{10}$(5-GHz&Ref.     \\
&Counts&Error&Counts&Error&density (nJy)&lum.\ density,&unres.&density(mJy)&lum.\ density,\\
&&&&&&W Hz$^{-1}$ sr$^{-1}$)&size (kpc)&&W Hz$^{-1}$ sr$^{-1}$)\\
\hline
3C\,318&$<$21.8 &-- &$<$21.8 &-- &$<$55.0 &$<$ 19.26&130 &$<$90 &$<$ 25.28&1 \\
3C\,324&52 &23 &$<$52 &26 &$<$6.6 &$<$ 18.86&35 &$<$0.14 $^*$&$<$ 22.91&3 \\
3C\,326&17 &7 &17 &7 &3.6 & 16.04&30 &13 & 22.56&1 \\
3C\,325&84 &25 &84 &25 &21 & 18.99&31 &2.4 & 23.83&5 \\
3C\,330&$<$19.2 &-- &$<$19.2 &-- &$<$13.0 &$<$ 18.30&110 &0.74 & 22.90&5 \\
NGC\,6109&85 &28 &59 &13 &12 & 15.57&2.5 &28 & 21.92&1 \\
3C\,334&1046 &41 &1075 &38 &342.0 & 19.73&26 &111 & 25.09&4 \\
3C\,338&3515.6 &76.6 &59.23 &22.06 &17.50 & 15.74&2.5 &105 & 22.50&1 \\
3C\,343&$<$8.7 &-- &$<$8.7 &-- &$<$4.6 &$<$ 18.48&140 &$<$300 &$<$ 26.05&21 \\
3C\,343.1&$<$22.3 &-- &$<$22.3 &-- &$<$12.4 &$<$ 18.61&130 &$<$200 &$<$ 25.62&1 \\
NGC\,6251&670 &33 &660 &33 &369 & 16.87&2.0 &850 & 23.23&24 \\
3C\,346&1130 &60 &767 &36 &155 & 18.20&49 &220 & 24.30&1 \\
3C\,345&1706 &47 &1706 &47 &496.0 \dag& 19.96&110 &8610 \dag& 27.04&2 \\
3C\,351&1008 &39 &1008 &39 &156.0 & 18.99&89 &6.5 & 23.49&4 \\
3C\,356&14 &8 &$<$14 &8 &$<$3.1 &$<$ 18.40&34 &1.1 & 23.70&5 \\
4C\,16.49&47 &17 &47 &17 &48 & 19.80&35 &16 & 25.04&23 \\
3C\,368&23 &8 &$<$23 &8 &$<$3.8 &$<$ 18.54&150 &$<$0.14 $^*$&$<$ 22.85&3 \\
3C\,380&1109 &37 &1109 &37 &656.0 & 20.24&120 &7447 & 27.11&2 \\
3C\,382&2912.8 &55.6 &2912.8 &55.6 &5950.0 & 18.85&4.7 &188 & 23.33&1 \\
3C\,388&1936 &266 &99 &19 &18 & 16.74&7.0 &62 & 23.26&1 \\
3C\,390.3&950 &32.3 &950 &32.3 &4320 \dag& 18.70&4.6 &330 \dag& 23.56&1 \\
3C\,433&$<$2.7 &-- &$<$2.7 &-- &$<$13 &$<$ 16.71&33 &5 & 22.30&1 \\
3C\,442A&316 &119 &$<$14 &5 &$<$15 &$<$ 15.54&2.2 &2 & 20.70&1 \\
3C\,449&1840 &85 &33 &12 &16 & 15.20&6.3 &37 & 21.57&1 \\
3C\,454&$<$4 &-- &$<$4 &-- &$<$20 &$<$ 19.70&38 &$<$200 &$<$ 26.43&20 \\
3C\,454.3&7625 &91 &7625 &91 &1350 \dag& 20.79&130 &12200 \dag& 27.53&2 \\
3C\,455&$<$4 &-- &$<$4 &-- &$<$20 &$<$ 18.30&25 &1.4 $^*$& 23.18&19 \\
3C\,465&439 &40 &439 &40 &65.8 & 16.30&2.5 &270 & 22.90&1 \\
\end{tabular}
\vskip 10pt
\begin{minipage}{\linewidth}
$*$ 5-GHz flux density
extrapolated from other radio frequencies.\\ \dag Known variability in
the marked waveband.\\
Refer to Table \ref{sources} to determine the instrument from which
the counts in columns 2--5 are taken. Total counts in column 2 are the
net in the source region after background subtraction and include any
unresolved component. Columns 6 and 7 list the flux and spectral luminosity
density at 1 keV of the unresolved component; the luminosity density
values quoted in columns 7 and 10 are the logs to base 10 of the
rest-frame spectral luminosity density in W Hz$^{-1}$ sr$^{-1}$, calculated
assuming $\alpha_X = 0.8$ and $\alpha_R = 0$. Column 8 gives the
approximate radius, in kpc, of an object that would be unresolved with
the appropriate {\it ROSAT} detector at the redshift of the source,
based on the half-energy radii quoted in section \ref{analysis}.
References in column 11 give the source of the 5-GHz core flux density
in column 9, and are as follows: (1) Giovannini \etal\ (1988); (2)
Hough \& Readhead (1989); (3) Best, Longair \& R\"ottgering (1997);
(4) Bridle \etal\ (1994); (5) Fernini \etal\ (1997); (6) Leahy, Bridle
\& Strom (1998); (7) Akujor \etal\ (1991); (8) Laing \& Riley, in
preparation; (9) Leahy, J\"agers \& Pooley (1986); (10) Sanghera
\etal\ (1995); (11) O'Dea \& Owen (1985); (12) Noordam \& de Bruyn
(1982); (13) Reid \etal\ (1995); (14) Saripalli \etal\ (1997); (15)
Gavazzi, Perola \& Jaffe (1981); (16) Strom \etal\ (1990); (17) Perley
\& Taylor (1991); (18) Liu, Pooley \& Riley (1992); (19) Bogers \etal\
(1994); (20) Cawthorne \etal\ (1986); (21) Pearson \& Readhead (1988);
(22) From MERLIN data of Hardcastle \etal\ (1997) (23) From maps of
Lonsdale, Barthel \& Miley (1993) (24) Jones \etal\ (1986) (25) Owen
\& Ledlow (1997) (26) Worrall \etal\ (1994).
\end{minipage} \end{table*}

\begin{table*}
\caption{Correlation strengths  and results of linear regression for logarithmic flux-flux relationships}
\label{fcor}
\begin{tabular}{lrrrrrrrr}
Source type&Number&$Z$-value&Significance&Theil-Sen
slope&\multicolumn{2}{c}{Range}&Schmitt slope&Intercept\\
&&&(per cent)&&Low&High\\
\hline
FRI radio galaxies&20&   4.390&$>99.99$& 0.99& 0.79& 1.24& 0.64&-5.68\\[5pt]
FRII low-excitation and&34&   2.699&99.31& 1.06& 0.59& 1.74& 0.56&-5.87\\narrow-line radio galaxies\\
All low-excitation and&54&   6.188&$>99.99$& 0.85& 0.72& 1.01& 0.69&-6.13\\narrow-line radio galaxies\\
All FRII radio galaxies&41&   4.270&$>99.99$& 1.59& 1.22& 2.16& 1.03&-6.46\\[5pt]
FRII quasars&20&   2.510&98.79& 0.80& 0.58& 1.13& 0.78&-5.24\\
All quasars&28&   3.680&99.98& 0.55& 0.40& 0.73& 0.79&-5.45\\[5pt]
All sources&89&   8.104&$>99.99$& 0.95& 0.86& 1.04& 0.86&-6.19\\
\end{tabular}
\vskip 10pt
\begin{minipage}{\linewidth}
$Z$-values are derived from the modified Kendall's $\tau$ test as
implemented in {\sc asurv}. Under the null hypothesis $Z$ is
approximately normally distributed with mean 0 and variance 1. The
intercepts quoted are those derived from the Schmitt regression, as
the Theil-Sen regression does not allow a derivation of the
intercept. The 90 per cent confidence ranges quoted for the Theil-Sen
slope are derived from Monte Carlo simulations (see Appendix A)
roughly matching the data and are only approximate. The equivalent
confidence ranges for the Schmitt slope, not shown, would be similar
but larger, as discussed in Appendix A.
\end{minipage}
\end{table*}

\begin{table*}
\caption{Correlation strengths and results of linear regression for
logarithmic luminosity-luminosity relationships}
\label{lcor}
\begin{tabular}{lrrrrrrrr}
Source type&Number&$Z$-value&Significance&Theil-Sen
slope&\multicolumn{2}{c}{Range}&Schmitt slope&Intercept\\
&&&(per cent)&&Low&High\\
\hline
FRI radio galaxies&20&   3.584&99.97& 0.99& 0.77& 1.28& 1.16&-10.33\\[5pt]
FRII low-excitation and&34&   3.609&99.97& 1.40& 1.02& 2.12& 1.35&-15.04\\narrow-line radio galaxies\\
All low-excitation and&54&   5.453&$>99.99$& 1.20& 1.03& 1.38& 1.28&-13.52\\narrow-line radio galaxies\\
All FRII radio galaxies&41&   4.031&99.99& 1.64& 1.33& 2.07& 1.52&-18.82\\[5pt]
FRII quasars&20&   4.116&$>99.99$& 0.50& 0.32& 0.75& 0.69& 2.26\\
All quasars&28&   4.776&$>99.99$& 0.42& 0.28& 0.57& 0.57& 5.12\\[5pt]
All sources&89&   9.467&$>99.99$& 1.16& 1.07& 1.26& 1.30&-13.69\\
\end{tabular}
\vskip 10pt
\begin{minipage}{\linewidth}
$Z$-values are derived from the modified Kendall's $\tau$ test as
implemented in {\sc asurv}. Under the null hypothesis $Z$ is
approximately normally distributed with mean 0 and variance 1. The
intercepts quoted are those derived from the Schmitt regression, as
the Theil-Sen regression does not allow a derivation of the
intercept. The 90 per cent confidence ranges quoted for the Theil-Sen
slope are derived from Monte Carlo simulations (see Appendix A)
roughly matching the data and are only approximate. The equivalent
confidence ranges for the Schmitt slope, not shown, would be similar
but larger, as discussed in Appendix A.
\end{minipage}
\end{table*}

\begin{table*}
\caption{Extended luminosities around $z>0.3$ 3CRR quasars and FRII radio
galaxies with unambiguous extended emission}
\label{exlumin}
\begin{tabular}{lrlllrrlr}
Object&$z$&Radio galaxy or&$kT$&2--10 keV lum.&Emission
line&Radius of pressure&Ref.&X-ray pressure\\
&&quasar&(keV)&($\times 10^{37}$ W)&pressure
(Pa)&measurement (kpc)&&(Pa)\\
\hline
3C\,48&0.367&Q&4.9&2.6&\\
3C\,215&0.411&Q&4.0&1.3&$3.2 \times 10^{-12}$&29&1&$6.1 \times 10^{-12}$\\
3C\,220.1&0.61&RG&5.6&4.2\\
3C\,254&0.734&Q&7.7&13&$3.5 \times 10^{-11}$&27&2&$1.7 \times 10^{-11}$\\
3C\,275.1&0.557&Q&4.8&2.5&$1.8 \times 10^{-12}$&27&1&$8.6 \times 10^{-12}$\\
3C\,280&0.996&RG&5.0&2.8\\
3C\,295&0.4614&RG&8.0&14\\
3C\,334&0.555&Q&4.4&1.9&$1.1 \times 10^{-11}$&27&1&$8.6\times 10^{-12}$\\
\end{tabular}
\vskip 10pt
\begin{minipage}{\linewidth}
X-ray gas temperature and luminosity (columns 4 and 5) are estimated
from the temperature-luminosity relation as discussed in the text.
References to emission-line measurements (column 8) are as follows: (1)
Crawford \& Fabian (1989); (2) Forbes \etal\ (1990). Pressures in
column 9 are calculated from the best-fitting $\beta$ model using the
method of Birkinshaw \& Worrall (1993).
\end{minipage}
\end{table*}

\appendix
\renewcommand{\thetable}{A\arabic{table}}
\begin{table}
\caption{Effects of bias in the distribution and censoring of the
variables}
\label{bias}
\begin{center}
\begin{tabular}{ll}
Data bias&Result\\
\hline
$p_x < p_y$&Slope overestimated\\
$p_x > p_y$&Slope underestimated\\
$\sigma_x > \sigma_y$&Slope underestimated\\
$\sigma_x < \sigma_y$&Slope overestimated\\
\end{tabular}
\end{center}
\end{table}

\clearpage
\begin{figure*}
\begin{center}
\leavevmode
\epsfxsize 12cm
\epsfbox{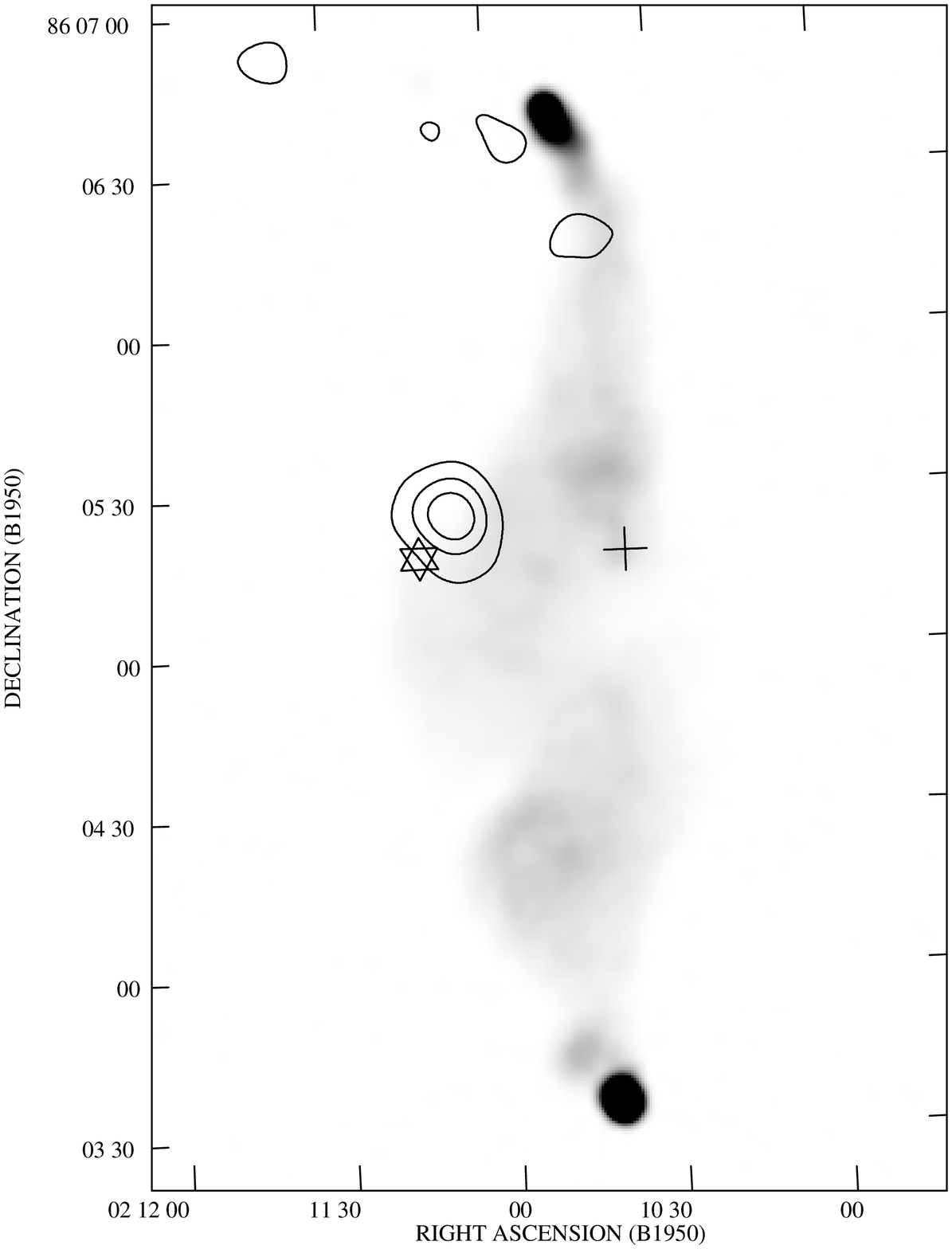}
\caption{3C\,61.1 in radio and X-ray. The greyscale is a 1.5-GHz radio
map (Leahy \& Perley 1991) taken from Leahy, Bridle \& Strom (1998);
the resolution is 3.7 arcseconds and black represents 100 mJy
beam$^{-1}$. Contours show the X-ray image smoothed with a $\sigma=4$
arcsec Gaussian, and are at $(6.1 \times 10^{-3}) \times (1, 2,
4\dots)$ HRI counts arcsec$^{-2}$. The lowest contour corresponds to
the $3\sigma$ level, calculated as described by Hardcastle \etal\
(1998c). The cross corresponds to the position of the 2.7-GHz radio
core (Laing \& Riley, in prep.) and the star to the position on DSS
plates of the background object discussed in the text. The offset
between the X-ray position and the position of the radio core is too
large to be introduced by {\it ROSAT} pointing inaccuracies.}
\label{61.1fig}
\end{center}
\end{figure*}
\begin{figure*}
\begin{center}
\leavevmode
\epsfxsize 12cm
\epsfbox{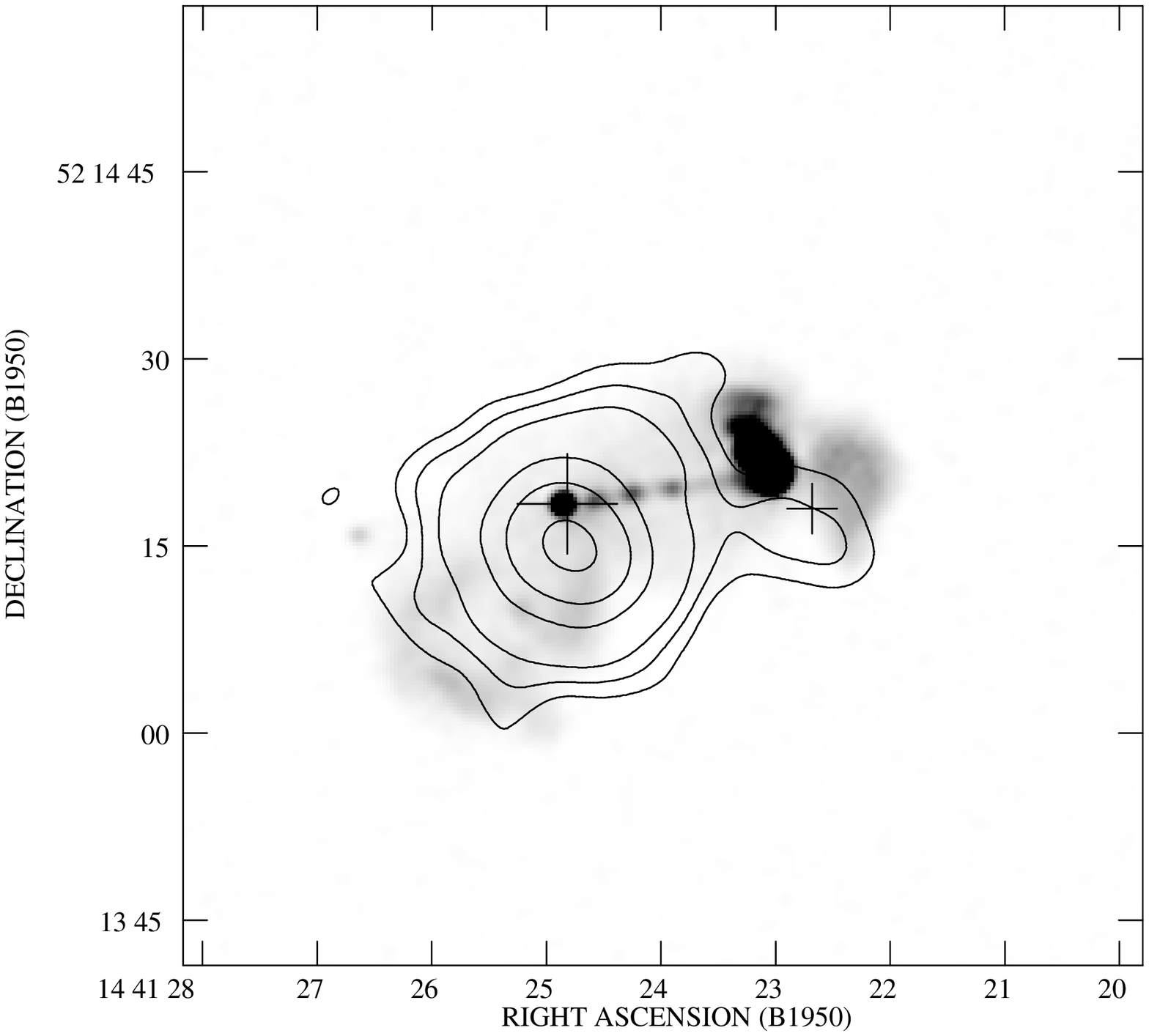}
\caption{3C\,303 in radio and X-ray. The greyscale is a 1.5-GHz radio
map (Leahy \& Perley 1991) taken from Leahy, Bridle \& Strom (1998);
the resolution is 1.2 arcseconds and black represents 20 mJy
beam$^{-1}$. Contours show the X-ray image smoothed with a $\sigma=2$
arcsec Gaussian, and are at $(8.03 \times 10^{-3}) \times (1, 2,
4\dots)$ HRI counts arcsec$^{-2}$. The lowest contour corresponds to
the $3\sigma$ level, calculated as described by Hardcastle \etal\
(1998c). The large cross corresponds to the optical position of the
host galaxy (Laing \& Riley, in prep.) and the smaller cross to the
optical position of the background quasar discussed in the text
(Kronberg 1976). The offsets between radio core and X-ray peak, and
between the western X-ray peak and either the radio hotspot or the
background quasar, are consistent with the absolute positional errors
in {\it ROSAT} data.}
\label{303-overlay}
\end{center}
\end{figure*}

\begin{figure*}
\begin{center}
\leavevmode
\epsfxsize 12cm
\epsfbox{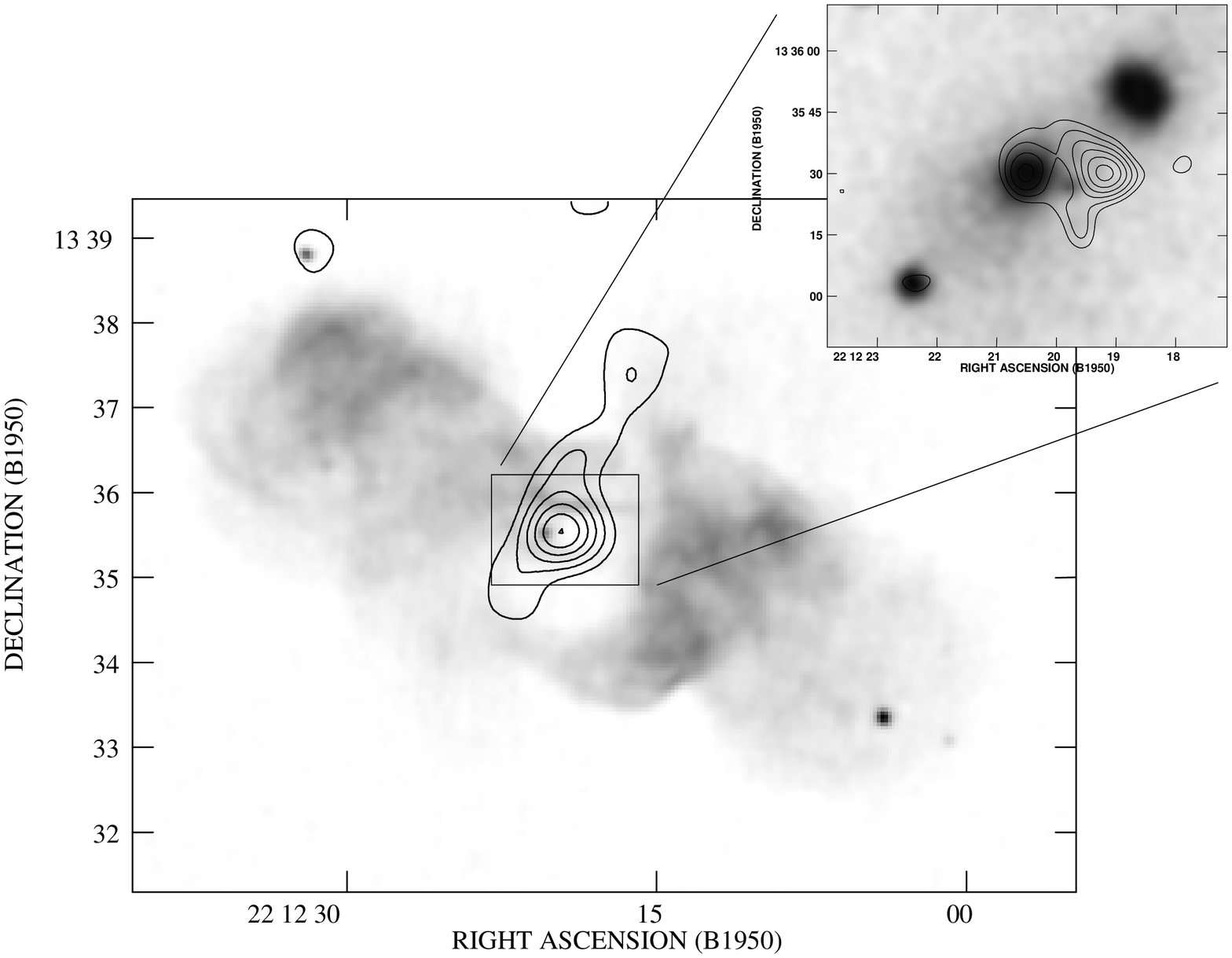}
\caption{3C\,442A in radio and X-ray. The greyscale is a 1.5-GHz radio
map (Comins \& Owen 1991) taken from Leahy, Bridle \& Strom (1998);
the resolution is 7.5 arcseconds and black represents 10 mJy
beam$^{-1}$. Contours show the X-ray image smoothed with a $\sigma=16$
arcsec Gaussian and are at $0.019 \times (1, 1.2,
1.4\dots)$ HRI counts arcsec$^{-2}$. The lowest contour corresponds to
the $3\sigma$ level, calculated as described by Hardcastle \etal\
(1998c). Inset is an overlay of the X-rays smoothed with a $\sigma=4$
arcsec Gaussian (contour levels at $0.044 \times (1, 1.2,
1.4\dots)$ HRI counts arcsec$^{-2}$, where the lowest level is again
$3\sigma$) on the optical image from the Digitised Sky Survey, showing
the apparent double X-ray structure. The central galaxy is NGC 7237,
that to the NW is NGC 7236.}
\label{442a-overlay}
\end{center}
\end{figure*}

\begin{figure*}
\begin{center}
\leavevmode
\epsfbox{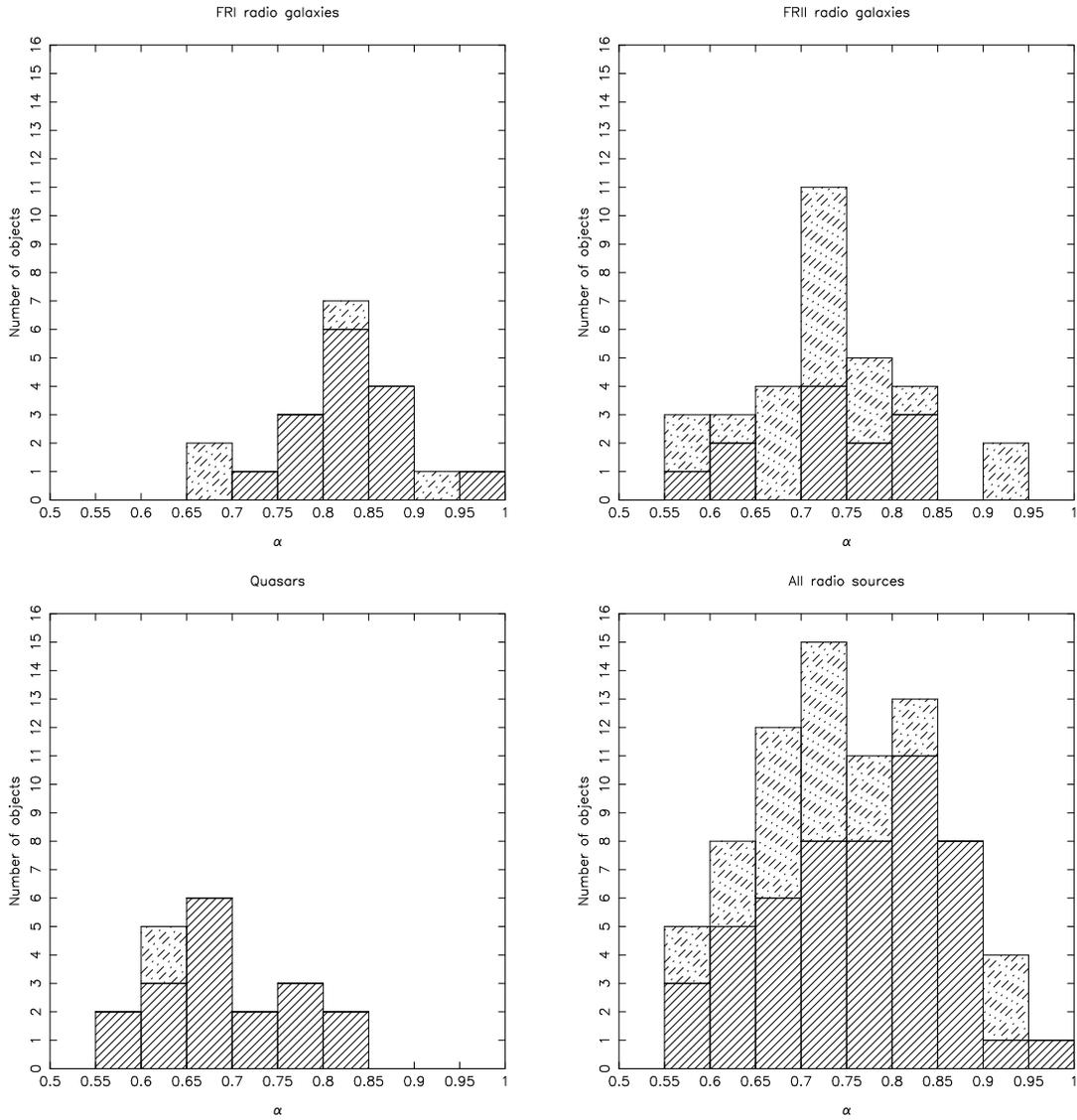}
\caption{Two-point rest-frame 5 GHz to 1 keV radio-X-ray core spectral indices for sources with detected radio
cores. Lightly shaded bins occur where there is a radio detection but
no X-ray detection and indicate lower limits on the spectral index.}
\label{alpha}
\end{center}
\end{figure*}

\begin{figure*}
\begin{center}
\leavevmode
\epsfbox{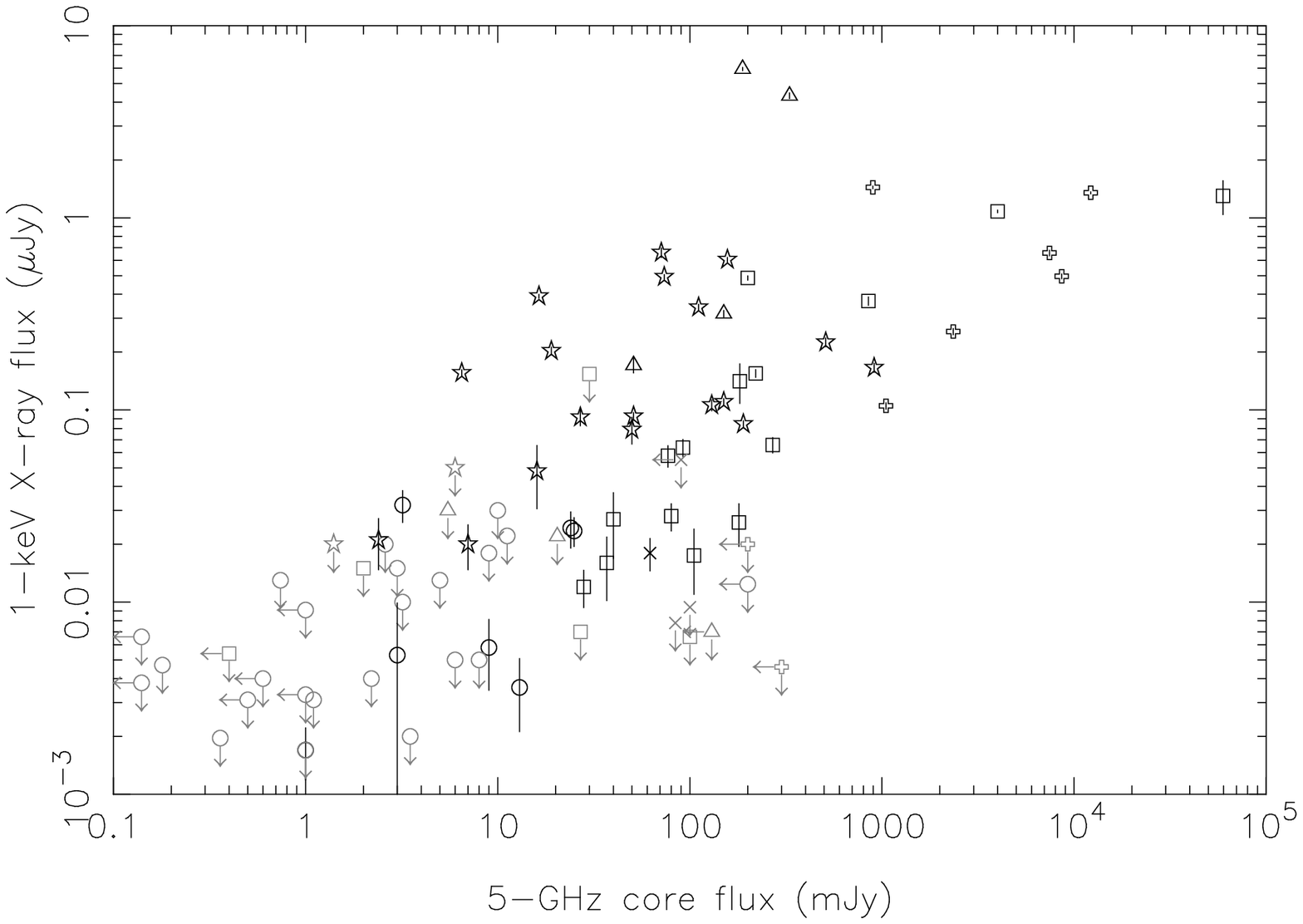}
\caption{X-ray core vs. 5-GHz radio core flux density. Symbols are as
follows; box, FRI radio galaxy; X, low-excitation FRII radio galaxy;
circle, narrow-line FRII radio galaxy; triangle, broad-line FRII radio
galaxy; star, lobe-dominated (FRII) quasar; cross,
core-dominated/compact quasar. Arrows denote limits; points with one
or more limits are plotted more lightly so as to make detections
easier to see.}
\label{ff}
\end{center}
\end{figure*}

\begin{figure*}
\begin{center}
\leavevmode
\epsfbox{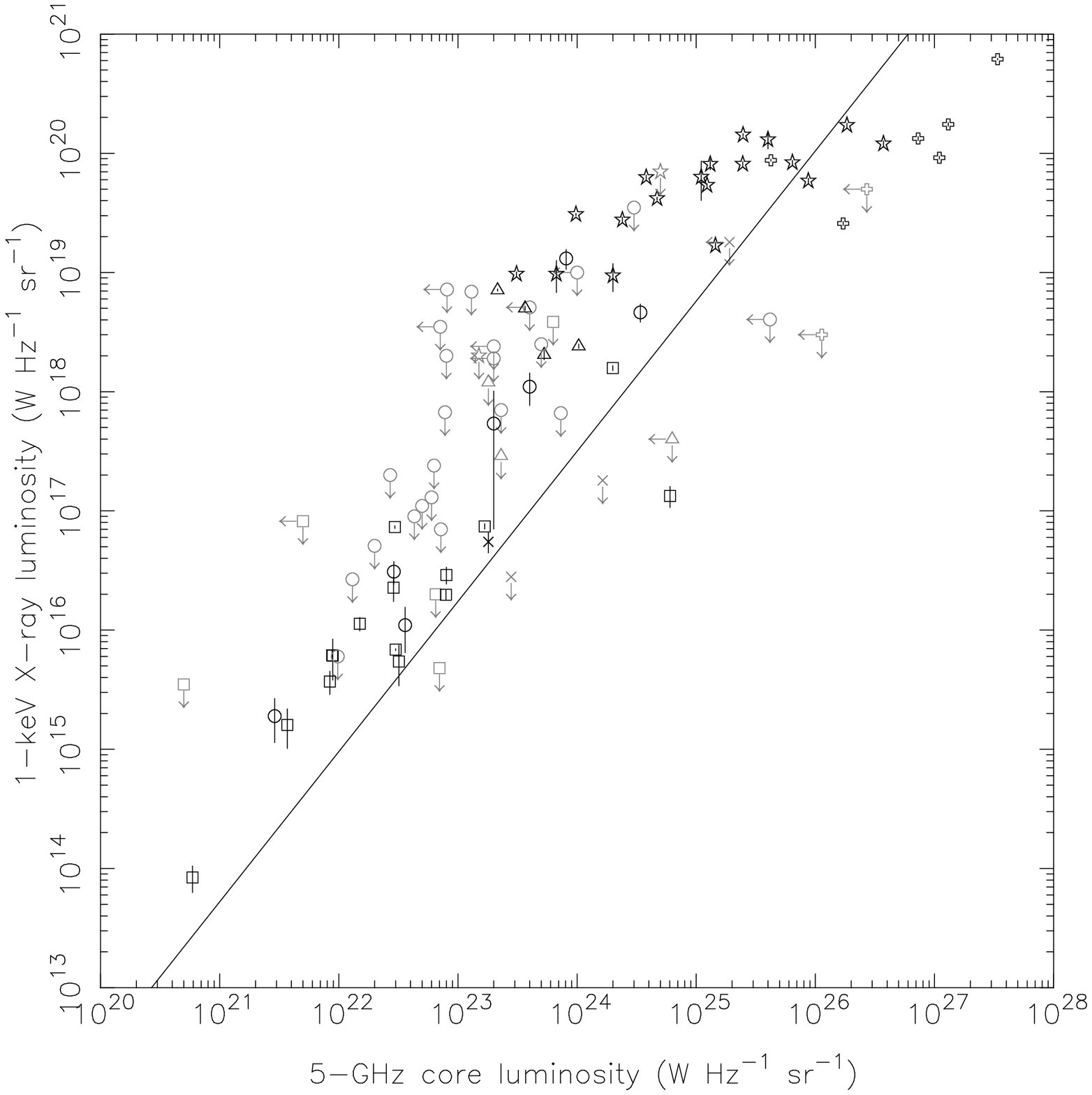}
\caption{X-ray core luminosity vs. 5-GHz radio core luminosity. Symbols are as
follows; box, FRI radio galaxy; X, low-excitation FRII radio galaxy;
circle, narrow-line FRII radio galaxy; triangle, broad-line FRII radio
galaxy; star, lobe-dominated (FRII) quasar; cross,
core-dominated/compact quasar. Arrows denote limits; points with one
or more limits are plotted more lightly so as to make detections
easier to see. The solid line shows the result of a
Schmitt linear regression for the whole sample (see Table
\ref{lcor}).}
\label{ll}
\end{center}
\end{figure*}

\begin{figure*}
\begin{center}
\leavevmode
\epsfbox{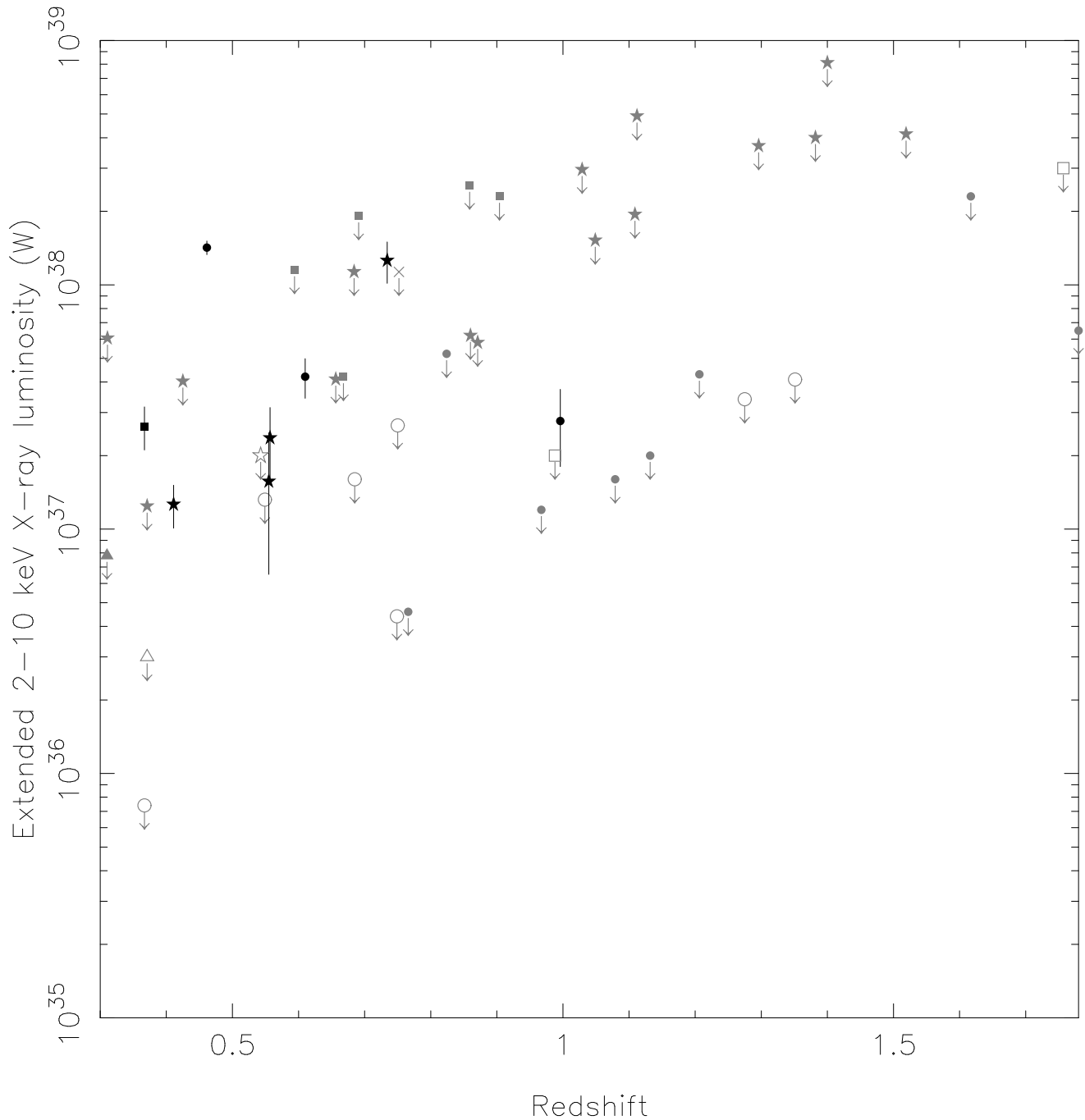}
\caption{Extended X-ray luminosity vs. redshift for high-redshift
objects. Symbols are as follows; circle, narrow-line FRII radio
galaxy; triangle, broad-line FRII radio galaxy; star, lobe-dominated
(FRII) quasar; box, core-dominated/compact quasar. Arrows denote
limits; points with one or more limits are plotted more lightly so as
to make detections easier to see. A filled symbol denotes a source
which is detected in the X-ray; open symbols are undetected. Only
sources where a separation beween pointlike and extended emission has
been made are plotted as detections; others are plotted as upper
limits. Where a source appears entirely pointlike, simulations (as
discussed in the text) are used to calculate an upper limit on the
luminosity of an undetected extended component.}
\label{extz}
\end{center}
\end{figure*}

\end{itemize}

\end{document}